\documentclass[a4paper,fleqn,usenatbib]{mnras}
\usepackage[T1]{fontenc}
\usepackage{ae,aecompl}

\usepackage{graphicx}	% Including figure files
\usepackage{amsmath}	% Advanced maths commands
\usepackage{amssymb}	% Extra maths symbols
\usepackage[dvipsnames]{xcolor}
\graphicspath{{figures/}}
\usepackage{hyperref}

\newcommand{\lya}{Lyman-$\alpha$}

\newcommand{\skm}{\mathrm{km^{-1}\,s}}

\newcommand{\ditto}{{\tt "}}

\newcommand{\be}{\begin{equation}}
\newcommand{\ee}{\end{equation}}

\def\orcid#1{\href{https://orcid.org/#1}{\includegraphics[keepaspectratio,width=0.7em]{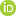}}}

\newcommand{\orcidauthorA}{\orcid{0000-0002-5445-461X}} % Irsic
\newcommand{\orcidauthorB}{\orcid{0000-0002-2642-5707}} % Viel
\newcommand{\orcidauthorC}{\orcid{0000-0001-8443-2393}} % Haehnelt
\newcommand{\orcidauthorD}{\orcid{0000-0003-2764-8248}} % Bolton
\newcommand{\orcidauthorF}{\orcid{0000-0001-8778-7587}} % Puchwein
\smallskip
\title[Ringing of the Reionization]
{Ringing of the Reionization: A first direct measurement of the intergalactic pressure smoothing scale at redshift z>4.2 as imprinted onto small-scale peculiar velocities in the Lyman-alpha forest}

\author[Ir\v{s}i\v{c} et al.]{Vid Ir\v{s}i\v{c}$^{1,2,3}$\,\orcidauthorA,\thanks{E-mail: v.irsic@hert.ac.uk (VI)}
Matteo Viel$^{4,5,6,7,8}$\,\orcidauthorB,
Martin G. Haehnelt$^{3,9}$\,\orcidauthorC, 
James S. Bolton$^{10}$\,\orcidauthorD, \newauthor
Ewald Puchwein$^{11}$\,\orcidauthorF,
Luke I. Gilmartin$^{12}$
\\
$^{1}$Center for Astrophysics Research, University of Hertfordshire, College Lane, Hatfield AL10 9AB, UK\\
$^{2}$Department of Physics, Astronomy and Mathematics, University of Hertfordshire, College Lane, Hatfield AL10 9AB, UK\\
$^{3}$Kavli Institute for Cosmology, University of Cambridge, Madingley Road, Cambridge CB3 0HA, UK\\
$^{4}$SISSA- International School for Advanced Studies, Via Bonomea 265, 34136 Trieste, Italy\\
$^{5}$INFN – National Institute for Nuclear Physics, Via Valerio 2, I-34127 Trieste, Italy\\
$^{6}$IFPU, Institute for Fundamental Physics of the Universe, via Beirut 2, 34151 Trieste, Italy\\
$^{7}$INAF, Osservatorio Astronomico di Trieste, Via G. B. Tiepolo 11, I-34131 Trieste, Italy\\
$^{8}$ICSC - Centro Nazionale di Ricerca in High Performance Computing, Big Data e Quantum Computing, Via Magnanelli 2, Bologna, Italy\\
$^{9}$Institute of Astronomy, University of Cambridge, Madingley Road, Cambridge CB3 0HA, UK\\
$^{10}$School of Physics and Astronomy, University of Nottingham, University Park, Nottingham, NG7 2RD, UK\\
$^{11}$Leibniz-Institut f\"ur Astrophysik Potsdam, An der Sternwarte 16, 14482 Potsdam, Germany\\
$^{12}$Lancaster University, Lancaster, LA1 4YB, UK
}
\date{Accepted XXX. Received YYY; in original form ZZZ}

\pubyear{2022}

\begin{document}
\label{firstpage}
\pagerange{\pageref{firstpage}--\pageref{lastpage}}
\maketitle

\begin{abstract}
The epoch of reionization leaves detectable imprints in the thermal history of the Universe. In particular, the fast ionization fronts passing over cold gas in the cosmic web overpressurize the gas, leading to the well-established phenomenon of pressure smoothing due to the hydrodynamic response of the gas. Although the effect has been indirectly measured from the power spectra of the Lyman-$\alpha$ forest over the past decades, very few examples exist of directly measuring the typical scale associated with this physical process. This work identifies an acoustic feature in the power spectrum of the projected peculiar velocity gradient $\eta$. Using toy models and linear theory, this work shows that the acoustic feature is likely strongly associated with the pressure smoothing scale. A feature at the same scale is imprinted on the small-scale flux power spectrum of the Lyman-$\alpha$ forest at $k=0.1-1\;\skm$. A methodology developed for the simulations is applied to the latest measurements of the flux power spectra at $z=4.2-5.0$ to provide the first direct measurements of the pressure smoothing scale at high redshifts, $\lambda_p(z=4.2)  > 33.78\;\mathrm{ckpc}\;(1\sigma)$, $\lambda_p(z=4.6) = 32.36\pm8.35\;\mathrm{ckpc}$ and $\lambda_p(z=5.0)=31.27\pm18.28\;\mathrm{ckpc}$. This proof-of-concept study paves the way for future observational programs aimed at recovering the thermal history using small-scale observations of the Lyman-$\alpha$ forest.
\end{abstract}


%%%%%%%%%%%%%%%%%%%%%%%%%%%%%%%%%%%%%%%%%%%%%%%%%%

%%%%%%%%%%%%%%%%% BODY OF PAPER %%%%%%%%%%%%%%%%%%

\section{Introduction}

The last phase transition of the Universe -- reionization -- provides a unique link between cosmology and astrophysics, connecting the onset of early galaxy formation and late time structure formation processes. As ionization fronts swept across the cosmic web they ionized and heated the neutral gas \citep{Miralda96,DAloisio2020}. The overpressurized gas expanded and cooled \citep{Gnedin98,Puchwein2023}, smoothing the gas density structure below a typical pressure smoothing scale \citep{kulkarni15}.

Reionization imprints signatures that are directly observable in the Cosmic Microwave Background \citep{Planck2020}, the \lya\, transmission of high redshift quasars \citep{Becker2015,Bosman2021} and \lya\, damping wings \citep{Eilers2017,Keating2024}; and others that are observable only indirectly through the thermal history of the intergalactic medium (IGM) \citep{Becker13,Boera2019,Gaikwad20}. Together, these measurements suggest a likely end to reionization at around $z\approx 5.3$ \citep{kulkarni19,Keating2019,Nasir2020,Bosman2021}. 

Due to the relatively long adiabatic cooling timescale, relics of reionization process persist into the IGM \citep{Hui97,Upton15,Hirata2018,Puchwein2023} deep into the post-reionization epoch, and their effect can be measured through the gas temperature evolution \citep{Becker13,Walther2018,Gaikwad21,Wilson2022} and the effects of the gas pressure smoothing scale \citep{Walther2018,Boera2019}. The pressure smoothing scale  is sensitive to the cumulative heat injection \citep{Theuns98,Gnedin98,Nasir16,Peeples10a,Peeples10b,Puchwein2023} and this provides an integral constraint over the periods of reionization that are unobservable with traditional IGM methods.

These methods that measure cumulative injected heat have typically been observed through an indirect inference using the Lyman-$\alpha$ forest -- a collection of absorption features due to background quasar light scattering off of neutral hydrogen in the IGM. These methods rely on comparison of the observations of the Lyman-$\alpha$ forest flux power spectrum and state-of-the-art, high-resolution hydrodynamical simulations (e.g. \cite{bolton17,palanque20}) to map observations onto the parameter spaces spanned by these simulations. An alternative method using the phase power spectrum of close quasar pairs was suggested by \citet{rorai13} and subsequently used to provide the first direct measurements of the pressure smoothing scale at redshift $2.0\leq z \leq 3.6$ \citep{Rorai2017}.

Aside from direct application to constraining reionization, the hydrodynamic response and the pressure smoothing it causes are also tightly linked to the understanding of the small-scale structure of the Lyman-$\alpha$ flux power spectrum, making it an important astrophysical caveat to address for the next generation of searches for the nature of dark matter (e.g. \cite{Rogers2021,Villasenor2022b,Irsic2024,GarciaGallego2025}). The impacts of the velocity and temperature fields during reionization on the small-scale Lyman-$\alpha$ forest have already been observed in previous works, highlighting the effects of self-shielding and relative baryon-dark matter velocity \citep{Cain2026}, the effects of temperature fluctuations \citep{Cain2024}, and peculiar velocity structure \citep{Molaro2022,Irsic2024}. This last aspect is what the current study builds upon.

The paper is structured as follows. Section~\ref{sec:methods} described the simulation setup used in this work, including the expanding spheres toy model in Section~\ref{sec:toy_model} and linear theory model in Section~\ref{sec:linear_theory}. Section~\ref{sec:results} presents the main results of the paper on simulations, followed by Section~\ref{sec:measurements} that presents the measurements on real data. The paper concludes with a review of the results against the literature in Section~\ref{sec:discussion} and with a summary in Section~\ref{sec:summary}.

\section{Methods}
\label{sec:methods}

\subsection{Simulations}
\label{sec:simulations}

This work is based on a suite of simulations from the Sherwood-Relics project \citep{bolton17,Puchwein2023}. These are high-resolution cosmological hydro-dynamical simulations suited for studying the high-redshift intergalactic medium. Full details on the simulations can be found in \citet{Puchwein2023,Irsic2024}, and are only summarized here for brevity.

The simulations were run with a customized version of {\tt P-Gadget3} \citep{springel05}, for different particle numbers, thermal histories, box sizes and cosmological models. This manuscript focuses on the results of cosmological boxes of size $40\,h^{-1}\;\mathrm{cMpc}$ with $2\times 1024^3$ dark matter and gas particles that extend down to $z=2$ (see Table~\ref{table:summary_sims}). The box size and resolution have been chosen to resolve the small scale structure that contributes to the flux power spectrum of the Lyman-$\alpha$ forest. For the purpose of numerical convergence tests,  further models were used with box sizes of $20$ and $10\;h^{-1}\;\mathrm{cMpc}$, and the same number of particles. All the models use a strongly simplified but very efficient star-formation prescription called {\tt Quick\_lya}, where gas particles are converted into collisionless star particles as they reach overdensities $\Delta > 10^3$ and temperatures $T < 10^5\;\mathrm{K}$ \citep{Viel04}. Simulations with different thermal histories were constructed using modifications of the spatially uniform UV background model by \citet{Puchwein19} using a non-equilibrium thermo-chemistry solver \citep{Puchwein15}. All simulated models investigated in this manuscript assume a flat $\Lambda$CDM cosmology with $\Omega_\Lambda = 0.692$, $\Omega_m = 0.308$, $\Omega_b = 0.0482$, $\sigma_8 = 0.829$, $n_s=0.961$, $h=0.678$ and $Y_p=0.24$ \citep{planck18}. The transmitted flux from the simulations is rescaled in post-processing \citep{lukic15,Bolton05} to match observations \citep{Boera2019}.

\begin{figure}
\centering
    \includegraphics[width=0.5\textwidth]{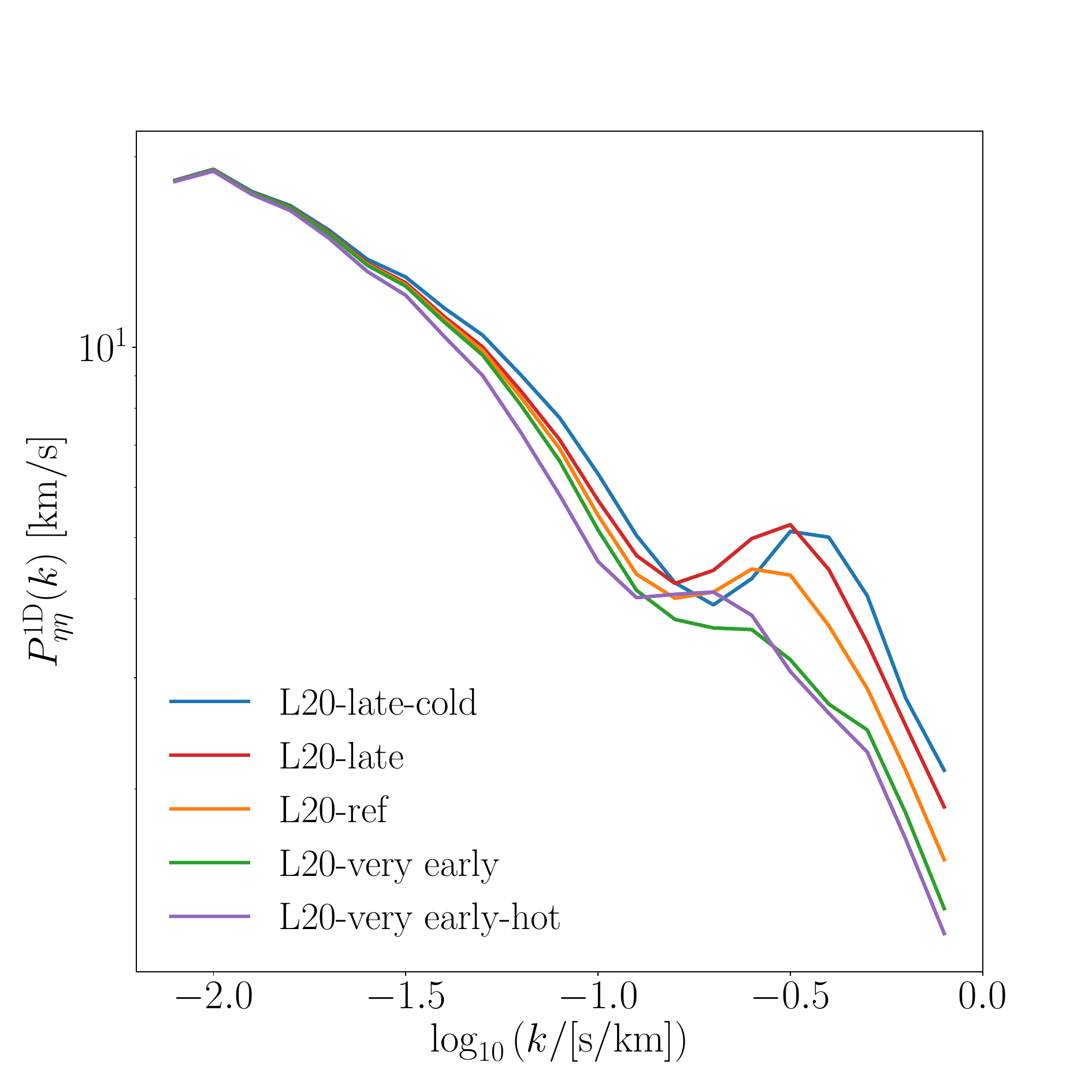}
\caption{The effect of varying thermal history on the effect of peculiar velocity power spectrum at $z=4.2$ in 20 $\mathrm{cMpc}/h$ box simulations (L20). The variations of the peculiar velocity fields are taken from different simulations of the thermal history, and are characterized by the cumulative injected heat ($u_0(4.2<z<12)$). 
%{\it Left:} 
The power spectrum of the velocity gradient ($\eta=-\nabla v_{\rm pec}/aH$) along the line-of-sight. The small-scale peculiar velocity structure is sourcing the effect in the flux power spectrum. The characteristic peak in the $\eta$ power spectrum suggests that the effect is coming from expansion velocities on small-scales of the order of $\sim 10\;\mathrm{km/s}$. These are the typical expansion velocities of the gas. The position of the peak also changes with respect to the $u_0$ value (see Table.~\ref{table:summary_sims}), indicating that the position of the peak is at smaller scales for models with lower cumulative injected heat. 
} \label{fig:models}
\end{figure}

The thermal history models are characterized by the cumulative injected heat parameter $u_0$ \citep{Nasir16,Boera2019} that captures the amount of heat injected into the IGM per baryon during the process of reionization. This parameter has a strong correlation with the pressure smoothing or filtering scale measured in simulations \citep{Gnedin98,kulkarni15,Boera2019}. The same simulated models can also be characterized by the temperature of the IGM at mean density $T_0$ and the redshift of the end of reionization $z_{\rm rei}^{\rm end}$ (when the volume averaged ionized fraction falls below $10^{-3}$). The exact conversions for different models are summarized in \citet{Puchwein2023,Irsic2024}. While the $u_0$ parameter is more closely related to the gas pressure smoothing, this manuscript uses $z_{\rm rei}^{\rm end}$ interchangeably, especially in the instances where it is important to highlight the difference between the time reionization ended and where the simulation outputs are being investigated. The simulations used in this work are summarized in Table.~\ref{table:summary_sims}. All units are in comoving space unless explicitly stated otherwise.

\begin{table}
\centering
\caption{List of simulations used in this work (see also \citep{Puchwein2023}). From left to right, the columns list the simulation name, the box size in $h^{-1}\rm\,cMpc$, the number of particles, the redshift of reionisation (defined as the redshift when the volume averaged ionised fraction $1 - x_{\rm HI} \leq 10^{-3}$) and the cumulative energy input per proton mass at the mean density, $u_{0}$, for $4.2\leq z \leq 12$ \citep[cf.][]{Boera2019}.
}
\label{table:summary_sims}
\begin{tabular}{lcccc}
\hline
Name     & $L_{\rm box}$ & $N_{\rm part}$ & $z_{\rm rei}^{\rm end}$ & $u_{0}(z=4.2)$ \\
 & $[h^{-1}\rm\,cMpc]$ & &  & $[\rm eV\,m_{\rm p}^{-1}]$ \\
\hline 
ref      & [10,20,40] & $2\times 1024^{3}$ & 6.00 & 8.15 \\
late & [20,40] & \ditto &  5.37 & 7.11 \\
early & \ditto & \ditto &  6.70 & 9.92 \\
very early & \ditto & \ditto &  7.40 & 11.61 \\
cold & \ditto & \ditto &  5.98 & 4.6 \\
late-cold & \ditto & \ditto &  5.35 & 4.02 \\
very early-hot & \ditto & \ditto &  7.41 & 21.1 \\
\hline
\end{tabular}
\end{table}

\subsection{Peculiar velocity field}

The primary focus of this work is on the small-scale clustering of the peculiar velocity gradient projected along the line of sight
\begin{equation}
\eta = - \frac{({\bf n}\cdot\nabla) ({\bf n}\cdot{\bf v})}{aH} = - \frac{\nabla_n v_{\rm pec}}{a H},    
\end{equation}
where ${\bf n}$ is the directional unit vector along the line of sight, ${\bf v}$ is the gas velocity and $H$ and $a$ are the Hubble rate and scale factor respectively. The projected peculiar velocity along the line of sight that gives rise to the redshift-space distortions is $v_{\rm pec} = {\bf n}.{\bf v}$, and $\nabla_n$ is the gradient in the direction of the line of sight. 

Because the observed flux transmission is related to the optical depth due to Lyman-$\alpha$ scattering in redshift space, the field $\eta$ is the first order redshift-space distortion correction \citep{Kaiser1987}. As summarized in Fig.~\ref{fig:models}, the 1D power spectrum of the $\eta$ field shows a distinct feature on scales of $0.1-1.0\skm$ in a variety of thermal history simulations (Table~\ref{table:summary_sims}). This scale is of the same order as expected from the pressure smoothing scale in the simulations \citep{Boera2019}, and leaves a distinct imprint on the flux power spectrum at a similar scale range, as first noted in \citet{Irsic2024}.

The feature clearly depends on the exact thermal history as shown in Fig.~\ref{fig:models}. While the models investigated in this study use a spatially homogeneous UV background, the inclusion of patchy reionization models does not significantly alter the picture \citep{Irsic2024}. The same feature as seen in the field $P_{\eta \eta}$ can also be observed directly in the power spectrum of the peculiar velocity field, rather than the gradient of the peculiar velocity. For that reason, the language used in this study often interchangeably uses both terms.

The feature appears on scales much smaller than typically investigated, and is thus susceptible to the issue of numerical convergence of the simulations themselves. While the numerical convergence of the flux power spectrum has been extensively studied, e.g. \citep{bolton17,Doughty2023}, including for the simulations used in this work \citep{Irsic2024}, the convergence of the peculiar velocity power spectrum is less well understood. A detailed comparison is presented in Appendix~\ref{appendix:0}. In general, the effect of numerical convergence on the velocity power spectra follows similar trends to that of the flux power spectra. While the exact amplitude and shape of the small-scale feature is affected by the numerical resolution, the position remains stable.

\subsection{Expanding spheres: a toy model}
\label{sec:toy_model}

\begin{figure}
\centering
\includegraphics[width=0.22\textwidth]{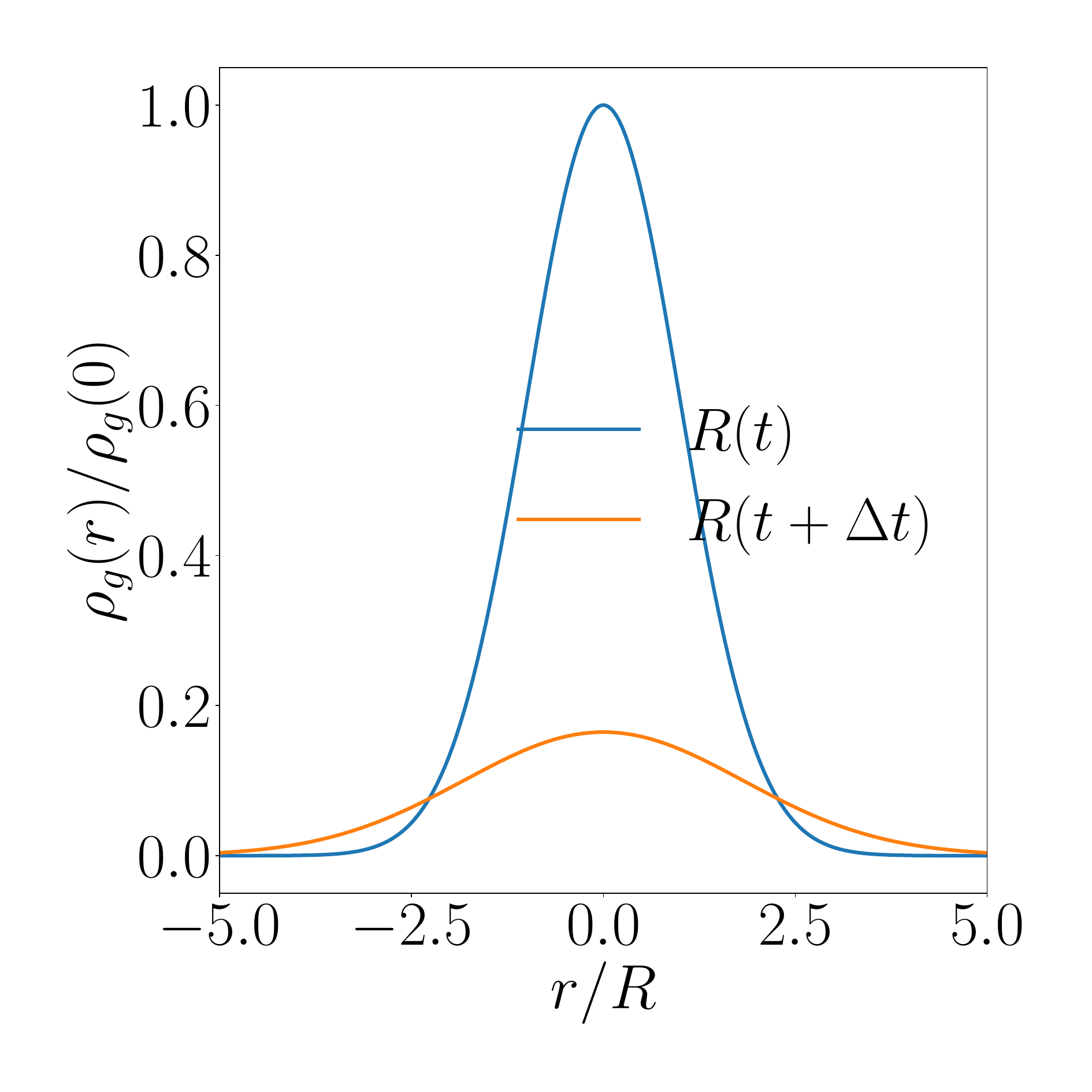} 
\includegraphics[width=0.22\textwidth]{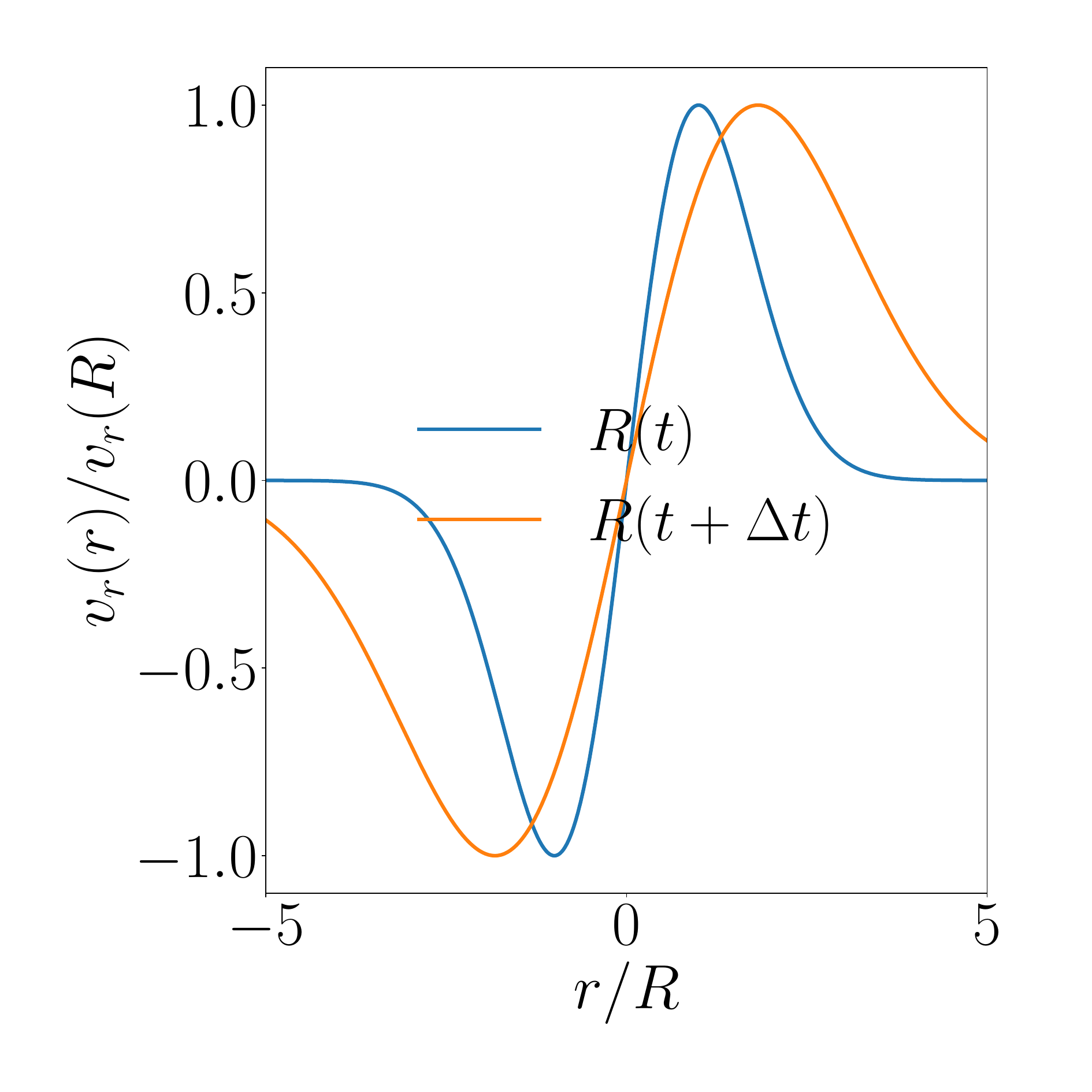} \\
    \includegraphics[width=0.22\textwidth]{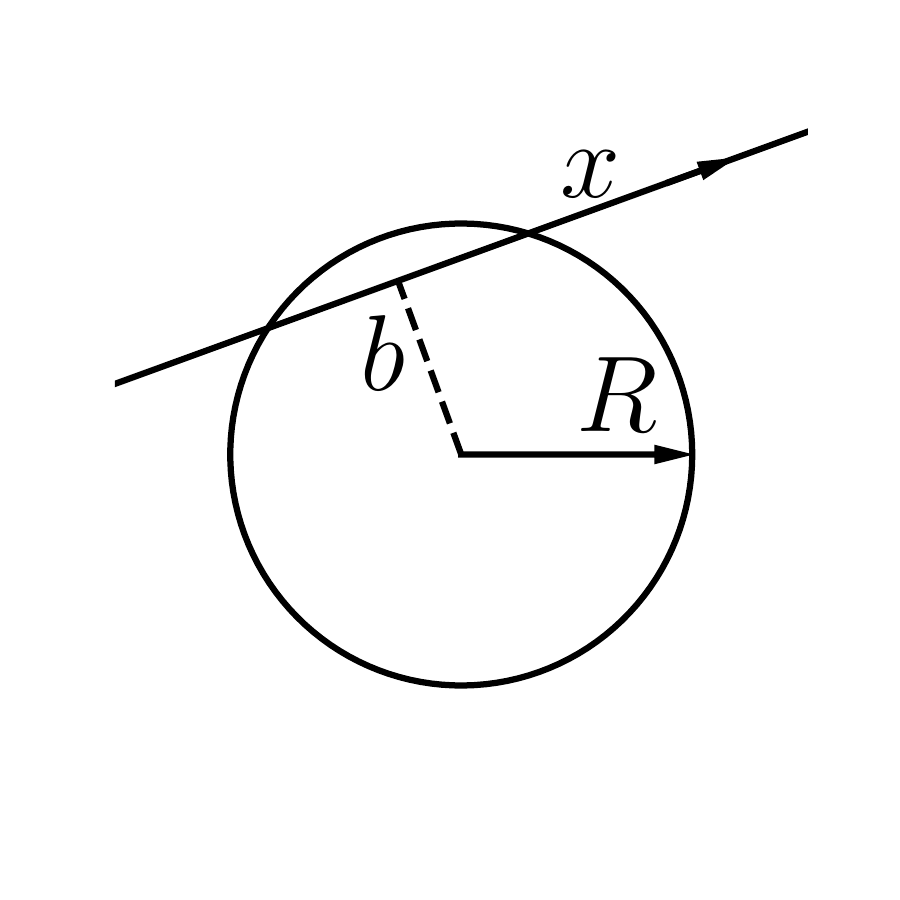}
    \includegraphics[width=0.22\textwidth]{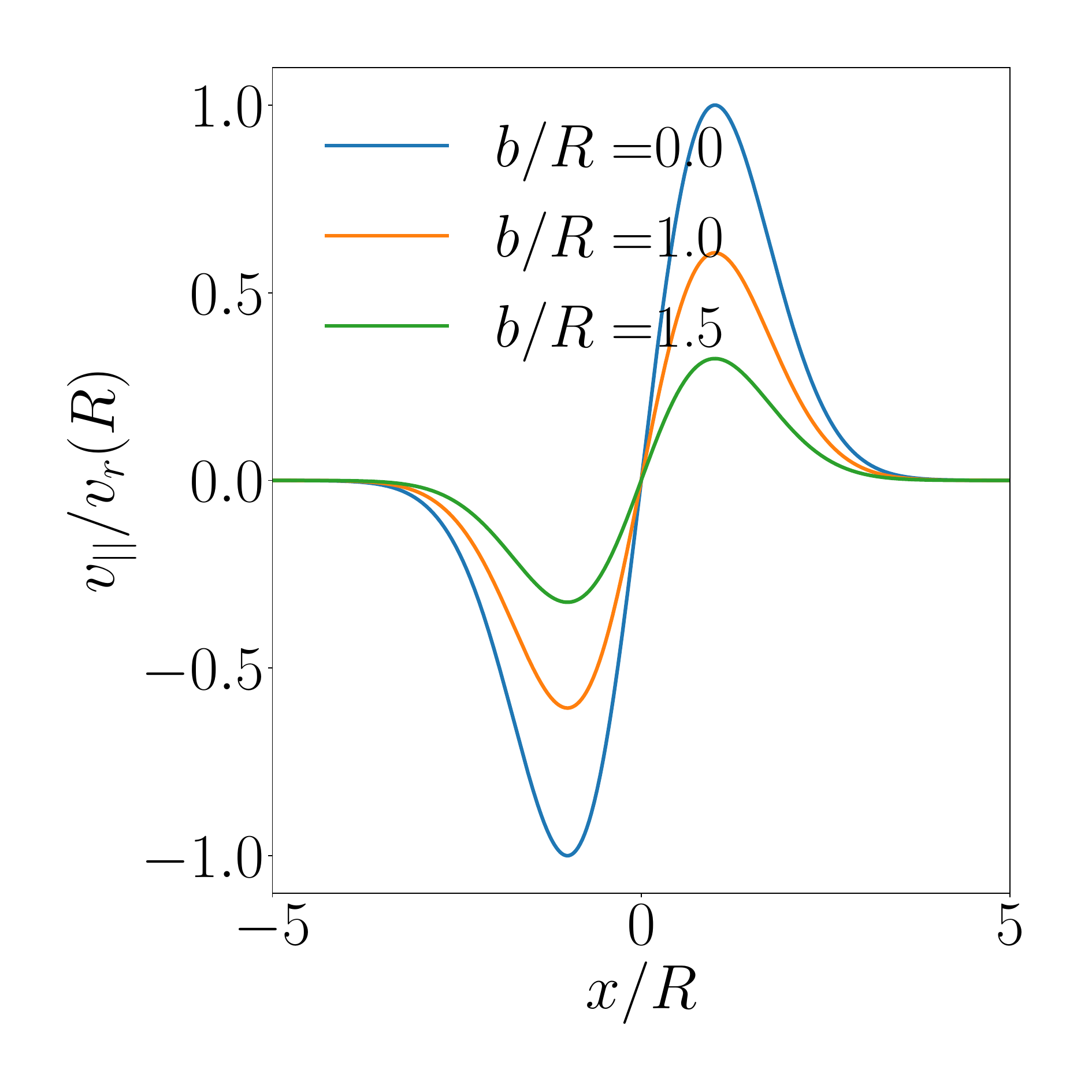}

\caption{A sketch of the density and velocity profiles around an expanding sphere in the toy model. The distances axes are scaled to the size of $R=R(t)$ at time $t$. {\it Bottom left}: the sketch of geometry, with the sphere radius at time $t$ given by $R$. The sphere is pierced by random sightline through the simulation box, at the impact parameter $b$. {\it Top left \& right:} The spherically symmetric density and peculiar velocity profile of the gas at two different times $t+\Delta t > t$. {\it Bottom right:} The projected peculiar velocity profile along the sightline piercing the sphere at different values of the impact parameter $b$.} \label{fig:sketch}
\end{figure}

A simple toy model is used to illustrate how the small-scale structure of the peculiar velocity field affects the flux power spectrum. The model includes a given number density of expanding spheres that are used as a proxy for the hydrodynamic response of the gas induced by the heat injection during reionization process.

To construct the model, the peculiar velocity field along each line-of-sight from L20-ref simulation is smoothed to obtain a baseline velocity field. The smoothed baseline field is constructed by convolving the original velocity field in the simulations with a Gaussian kernel. The smoothing scale for this process is chosen to be given by the thermal broadening scale, which was found to preserve the scale dependence at $k<0.1\;\skm$. A simple model of expanding spheres is then injected on top of the baseline field. 

The sphere positions are random throughout the volume of the simulation, given a comoving number density parameter ($\bar{n}$) with typical values of $\sim 10\;h^3\,\mathrm{cMpc^{-3}}$. This highlights the requirement for ubiquity of such objects to mimic the signal in simulations, with the number density comparable to that of $10^8-10^9\;h^{-1}\,\mathrm{M_\odot}$ halos. However, the expanding spheres in the model are not associated with halo positions, and the effect of clustering of the sphere positions has only a small effect on the peculiar velocity power spectrum.

The density profile of the spheres is modelled as a spherically symmetric Gaussian radial profile as shown in Fig.~\ref{fig:sketch}. The size, or standard deviation, of the profile is time dependent, which expands the sphere, and through mass conservation, also decreases the peak density in the centre of the sphere. Through the use of the continuity equation this leads to the peculiar velocity field being modelled as first order Hermite radial profile (see Appendix~\ref{appendix:B}). The characteristic size of the spherical density profile was given by free expansion, with the standard deviation parameter $R(t) = R + v \sqrt{2\pi e} (t-t_0)$, for a fixed expansion velocity $v$ and the initial size of the sphere given by $R=R(t_0)$. The numerical factor of $\sqrt{2\pi e}$ was chosen in such a way that the peak velocity of the radial peculiar velocity at radius $R(t)$ is exactly $v$.

This peculiar velocity field is then projected onto each sightline passing in its vicinity. While the domain support for a Gaussian (or Hermite) radial profile is infinite, in practice sightlines that are further away from the central position of the sphere than $3R$ contribute a negligible amount to the total velocity field along that sightline. All spheres in this toy model were injected with a constant value of $R$ and $v$.

\begin{figure*}
\centering
    \includegraphics[width=0.3\textwidth]{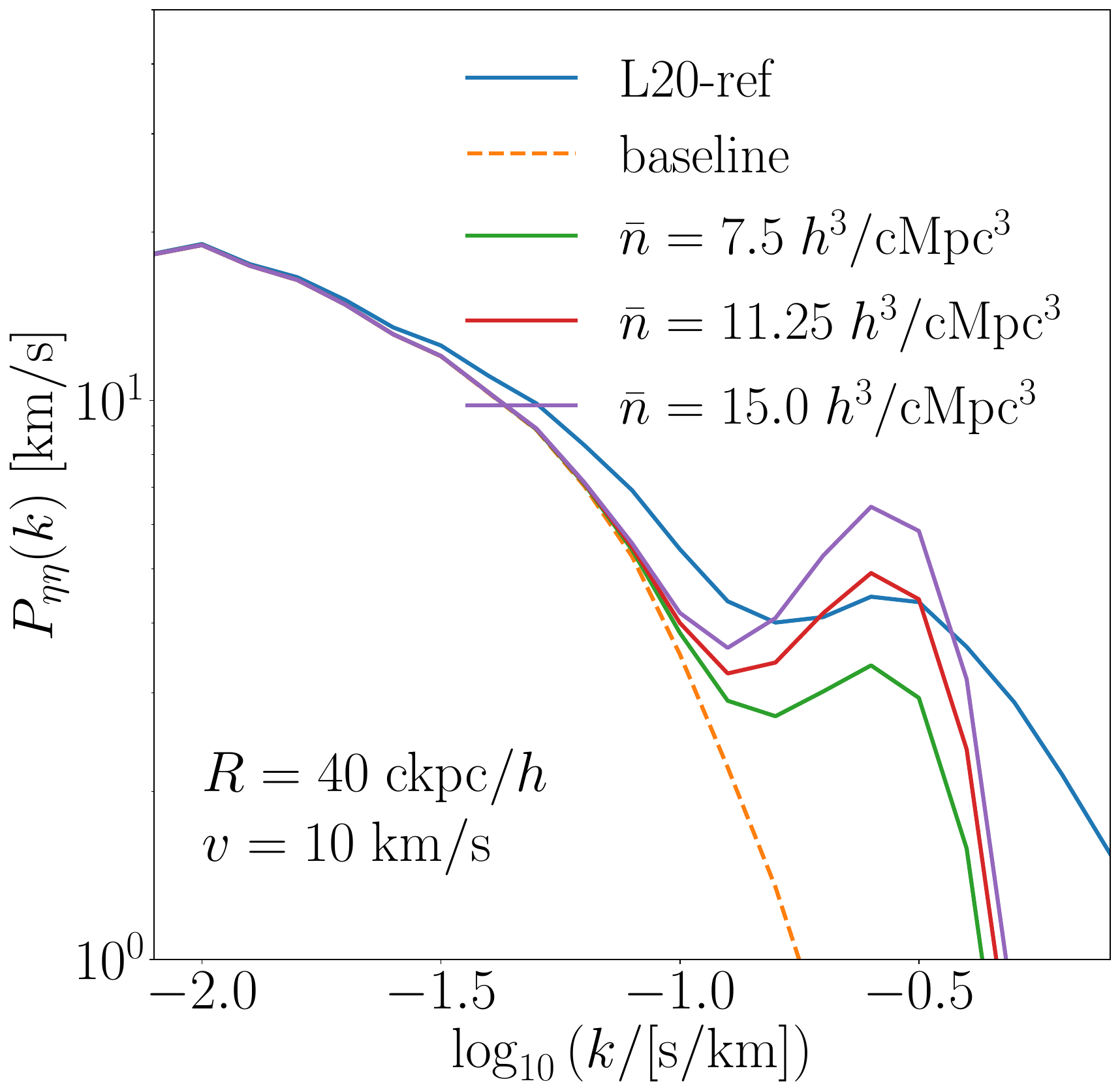}
    \includegraphics[width=0.3\textwidth]{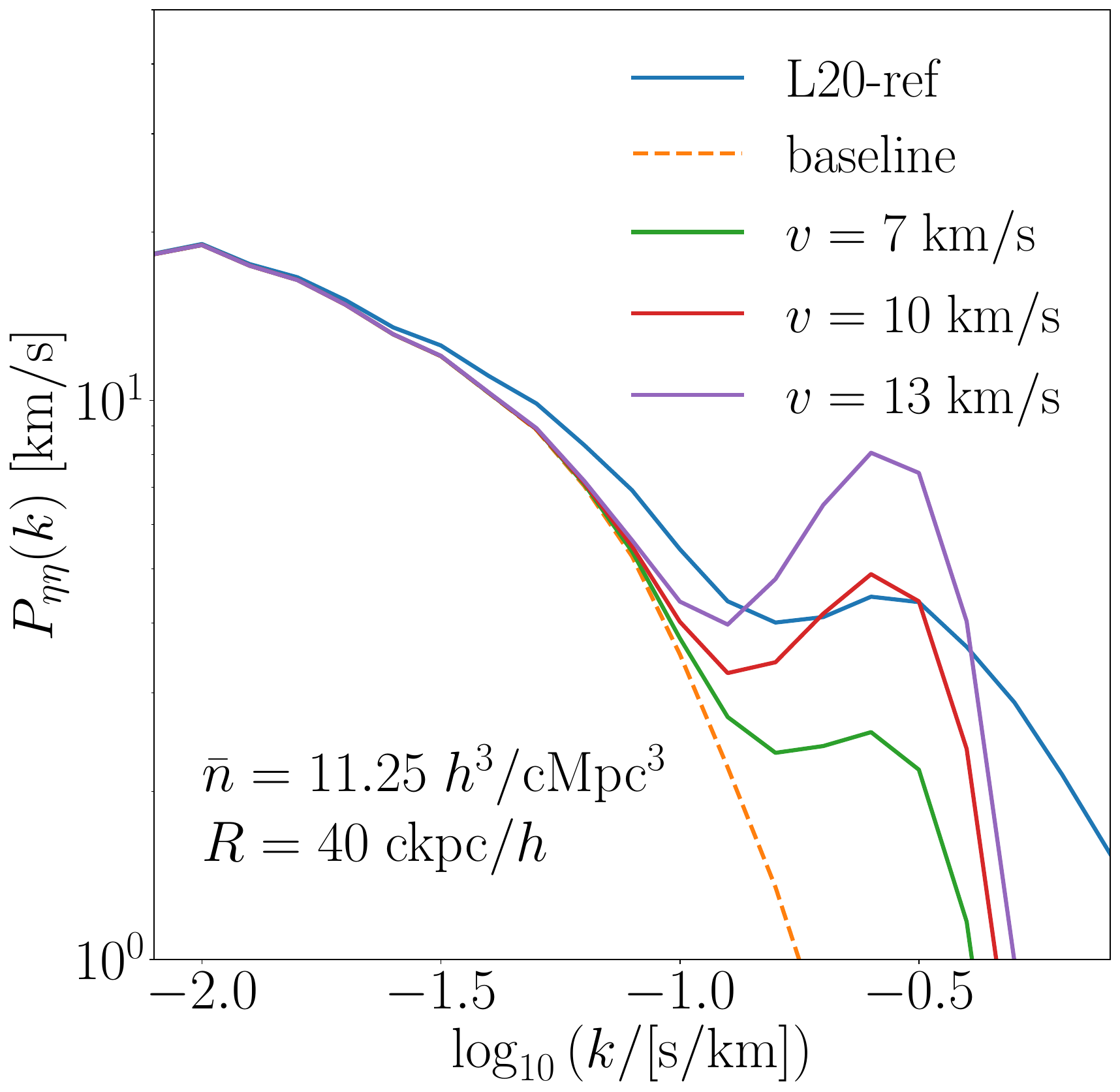}
    \includegraphics[width=0.3\textwidth]{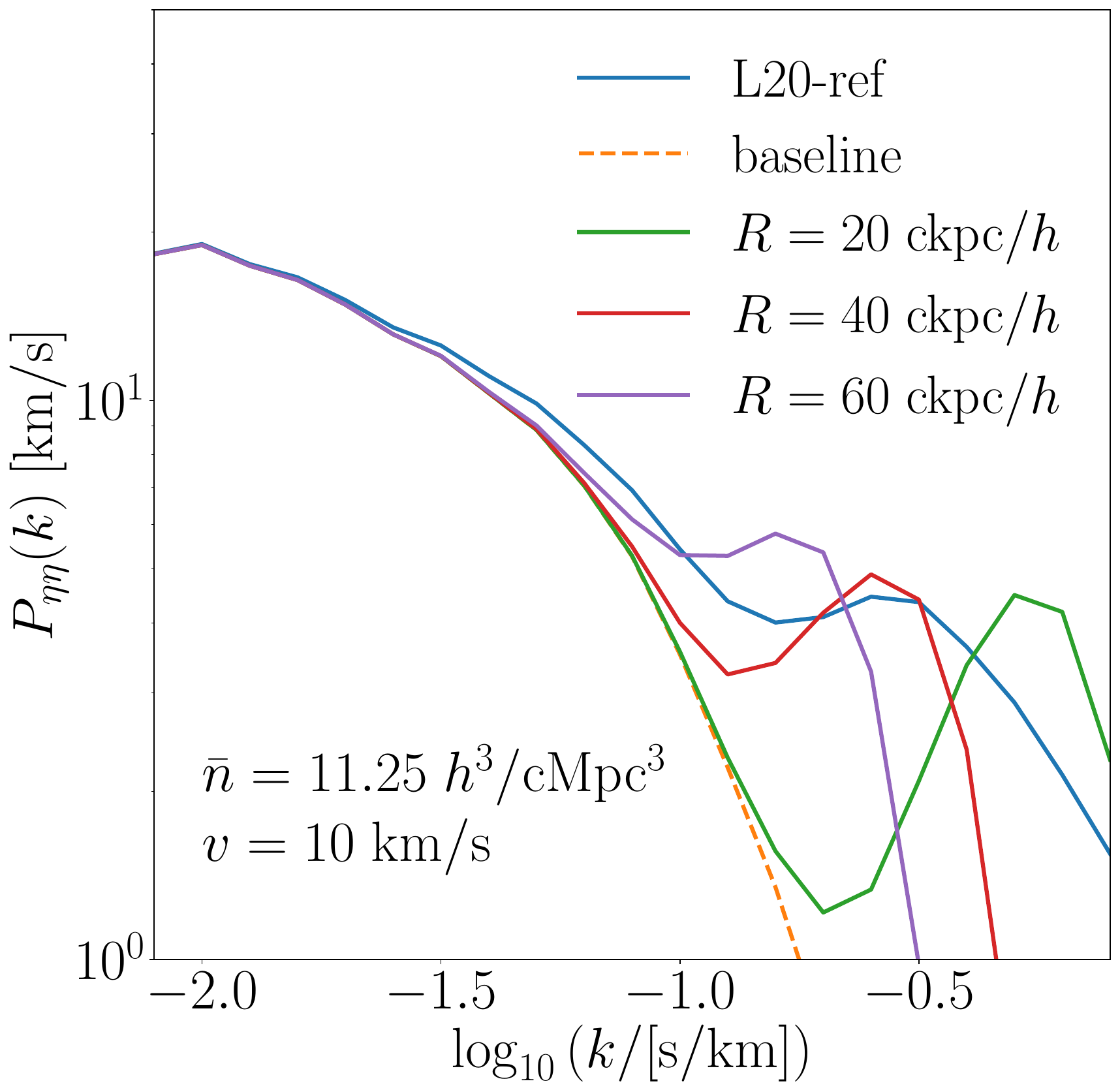}  \\

    \includegraphics[width=0.3\textwidth]{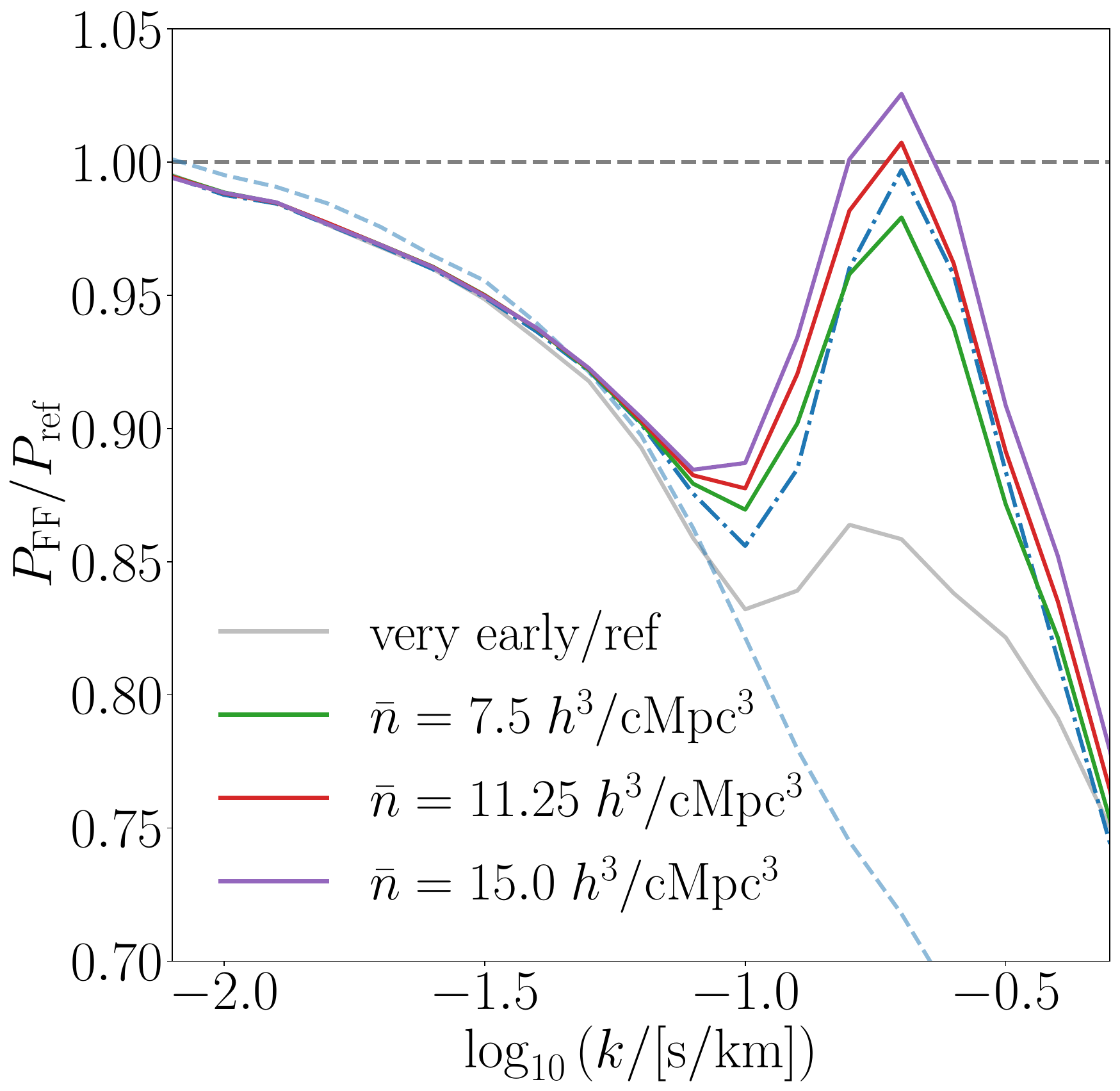}
    \includegraphics[width=0.3\textwidth]{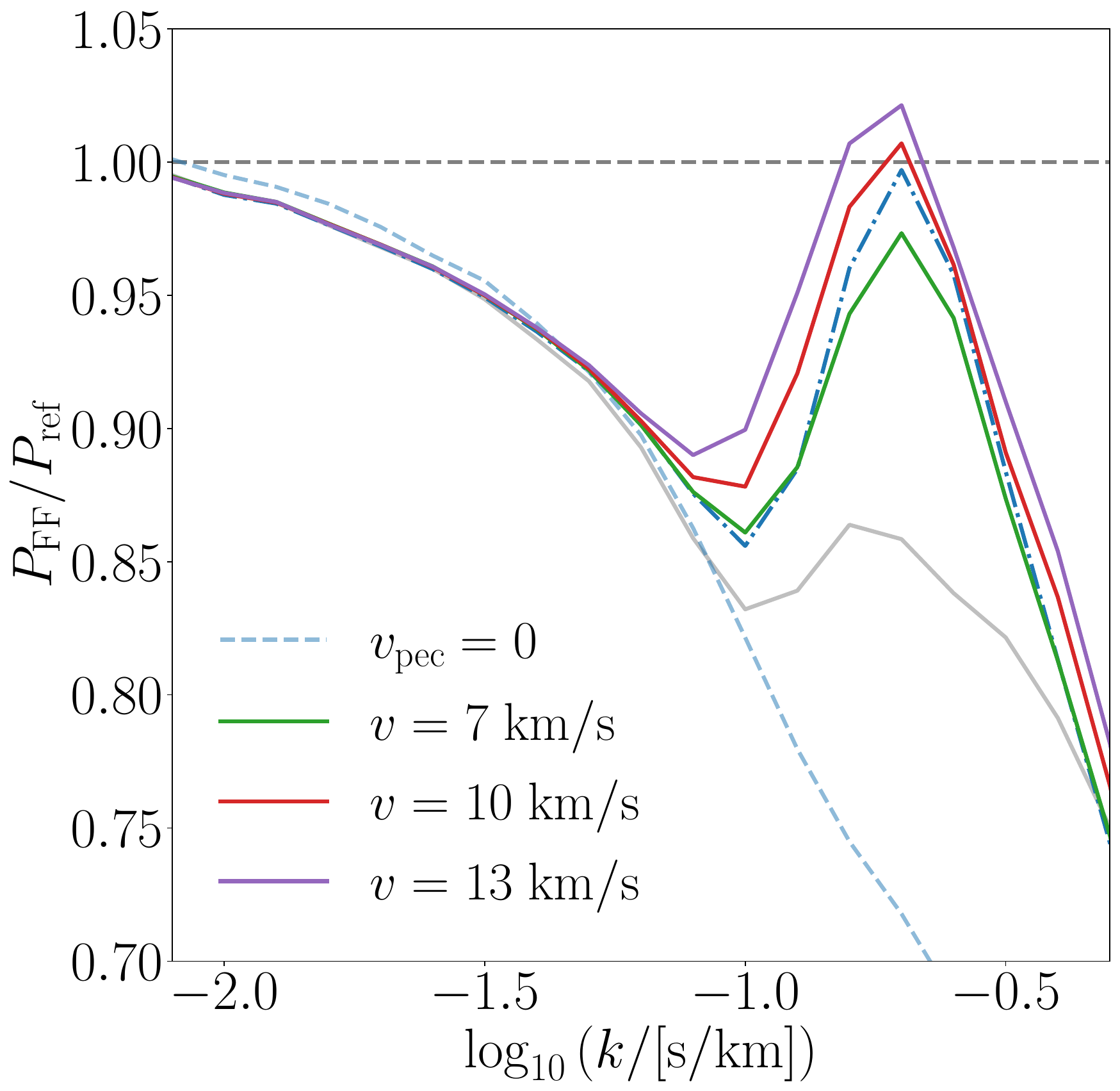}
    \includegraphics[width=0.3\textwidth]{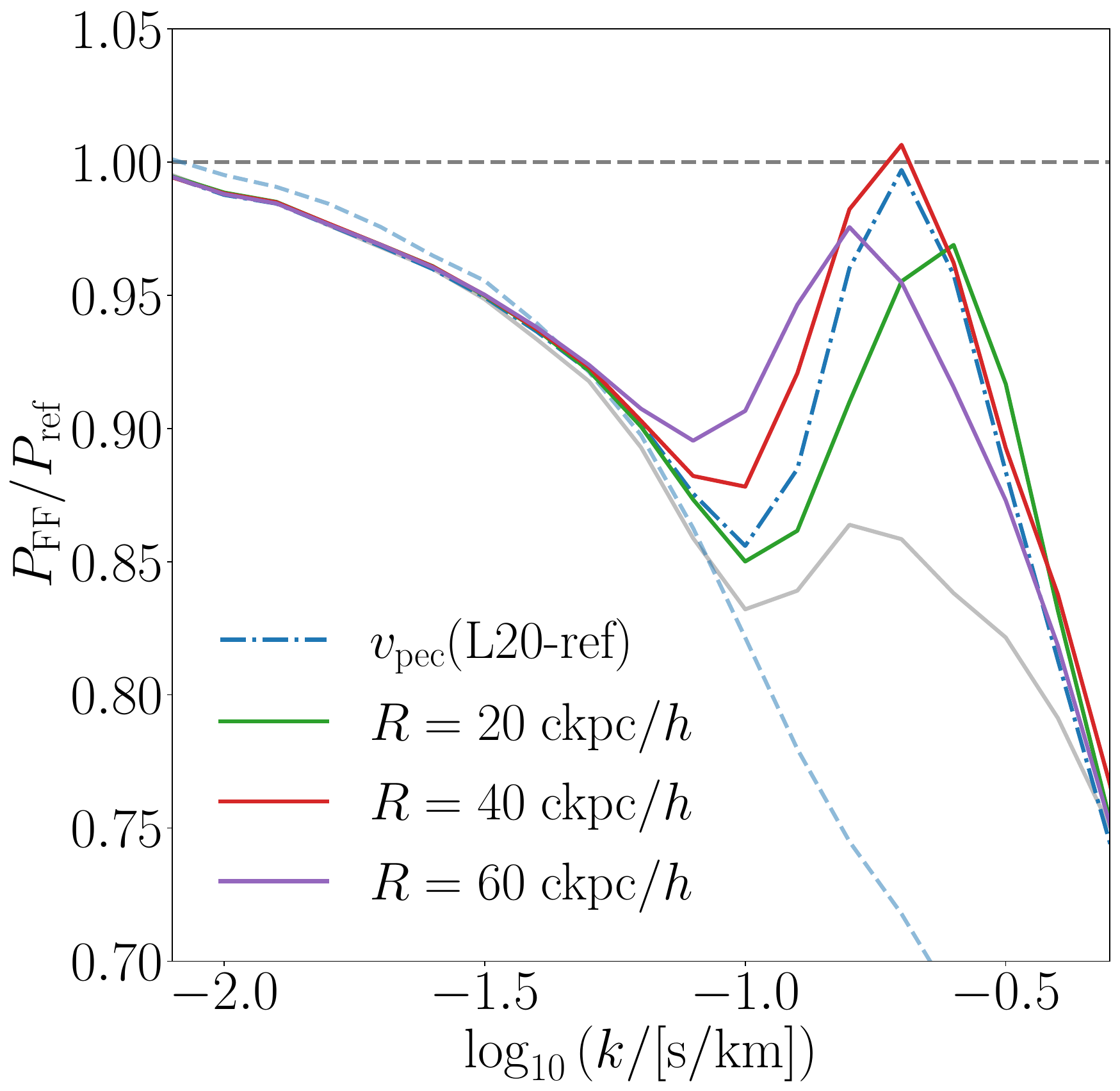}

\caption{The effects of varying parameters in the sphere model for the reference model (L20-ref) at $z=4.2$.
%\jamie{confusing to describe this now and then go on to explain the top row again} 
 Different columns correspond to model variations where only one parameter was varied at a time: number density of spheres ${\bar n}$ ({\it left}); the peak line-of-sight expansion velocity $v$ ({\it center}); and the proxy for the size of the spheres given as the typical scale of the radial profile $R$ ({\it right}). {\it Top:} The top row shows the velocity gradient power spectra. The sphere model was built on top of a baseline, which was taken to be a smoothed velocity field of the original simulation (dashed orange). The original simulations' peculiar velocity field power spectrum is shown in solid blue (L20-ref). {\it Bottom:} The ratio of the flux power spectrum between two models with different thermal history highlight the scales that are affected by peculiar velocity structure. This is compared to results where the optical depth to Lyman-$\alpha$ for both simulations was constructed with no peculiar velocities (dotted blue) or the same peculiar velocity field (dot-dashed blue) from the reference thermal history (L20-ref). The ratio of original simulations is shown in gray. 
} \label{fig:peta_model_params}
\end{figure*}

The effects of varying the input parameters of this toy model $(R,v,{\bar n})$ are shown in Fig.~\ref{fig:peta_model_params}. The top row of the figure shows how changing these parameters affects the peculiar velocity power spectrum, relative to the original simulation (solid blue) and the smooth baseline model (dotted orange). For typical parameter values of $({\bar n}=11.25 h^3/\mathrm{cMpc^3},v=10\;\mathrm{km/s},R=40\;\mathrm{ckpc/h})$ the small-scale structure in the toy model has a similar shape to that found in the original simulations. The relative amplitude of the small-scale peak can be modulated by increasing either the number density ${\bar n}$ or the peak velocity $v$, with slightly stronger sensitivity to the value of the velocity. The position of the small-scale structure, in this toy model, is solely determined by the size of the Gaussian density kernel of the sphere $R$. 

The toy model neglects that the expansion velocity of a sphere might depend on the local environment, and is likely dominated by the sound speed of the intergalactic gas which is determined by the temperature. Similarly, the size of the Gaussian kernel ($R$) likely depends on when the spheres started expanding, varying with both the environment and the patchy nature of reionization. The expansion of the spheres is likely also not a valid approximation in reality as the local sound speed velocity depends on the local density and might not be spherical symmetric, especially in the low density filamentary structure of the cosmic web that sources the Lyman-$\alpha$ forest. An example of such a structure in simulations was reported in \citep{Puchwein2023}. However, these dependencies are in reality highly non-trivial. For the sake of simplicity of the model, and to highlight the signal that expanding spheres create in the distribution of the peculiar velocity and flux fields along the lines-of-sight, we neglect the contribution of varying $R$ and $v$ across the box.

The bottom panels of Fig.~\ref{fig:peta_model_params} show the effect of varying toy model parameters on the flux power spectrum. The figures show how the effect drastically changes the relative small-scale clustering between two models with different thermal histories; a very early reionization (L20-very early) and the reference model (L20-ref) of \citep{Puchwein2023}. The ratio of flux power highlights three cases: (a) the original ratio of the flux power between the two simulations (in gray); (b)  the reconstructed flux power from both simulations when neglecting the effect of peculiar velocity field entirely $v_{\rm pec}=0$ (dashed blue); and (c) the reconstructed flux power where the same velocity field was used in the flux reconstruction, in this case the velocity field from the reference simulation (dashed blue). The difference between the first two cases shows that within simulations themselves the small-scale structure in the flux is driven by the peculiar velocity field, and the position of the small-scale structure coincides with that seen in the top panels in the peculiar velocity field. The last case is used as a reference point between simulations and the toy model, where the velocity field in both simulations is the same, either given by that of the reference simulation or constructed from the toy model. The flux power spectrum of the toy model shows similar behaviour as seen in the peculiar velocity power spectrum. Moreover, the same parameters that describe well the velocity power spectrum of the reference simulation, also describe the flux power spectrum ratio when both simulations are reconstructed with that same velocity field.

Despite the many assumptions of the toy model, it reproduces some of the key features of the hydrodynamical simulations. This is highlighted in Fig.~\ref{fig:peta_model_params} (bottom panels) where variations of the model parameters are indicative of the sensitivity to the more complex physics captured in the simulations. In particular, the model of expanding spheres reproduces the peak feature in the peculiar velocity power spectrum (Fig.~\ref{fig:peta_model_params};top), including its position and amplitude. The amplitude is largely determined by the expansion velocity of the spheres, and also modulated by their number density in the simulated volume. The number densities that reproduce simulations suggest the ubiquitous nature of the expanding structures in the simulation. The expansion velocities of $\sim 10\;\mathrm{km/s}$ further support the idea that the structures in the simulation hydrodynamically react to a temperature or pressure gradient and are expanding with the speed of sound in the environment. In this model, the position of the peak is determined solely by the size of the expanding spheres at the time of observation. The size of the structures of $\sim 40\;\mathrm{ckpc}/h$ would suggest that the structures were expanding for $\sim 1\;\mathrm{Gyr}$, if the expansion is dominated by the free-expansion with $v \sim 10\;\mathrm{km/s}$ \citep{Puchwein2023}.

While simplistic by design, the toy model presents an intuitive picture of the nature of this small-scale structure feature. It furthermore clearly links it to the structure of the peculiar velocity field on small-scales, rather than gas temperature or ionization fields. This agrees will with previous studies of simulations \citep{Molaro2022,Molaro2023}.

%%%%%%

\subsection{Linear theory}
\label{sec:linear_theory}

To further expand on the nature of the small-scale structure feature in the $P_{\eta\eta}$ power spectrum, the results of the simulations can be compared to the expectations from linear theory \citep{Peebles1980,Gnedin98,Barkana2001}. 

Using the coupled system of linearized equations for the growth of dark matter and baryon structure defines a characteristic filtering cut-off ($k_F$) in the baryon density \citep{Gnedin98} that arises due to the integrated pressure effect that links to thermal history through the global value of the sound speed. The suppression of density fluctuations also propagates to the peculiar velocity field, showing both the general filtering scale suppression and the signature of acoustic oscillations. 

The mapping between the peculiar velocity field and the density involves the time derivative $\eta =-\mu^2 a \delta_b'$, where $'=d/da$ and $\mu=k_{||}/k$. The transfer function that maps from the matter field to the gradient of the peculiar velocity $\eta$ is given as $T_\eta = a \delta_b'/\delta_c$, where $\delta_{c},\delta_b$ are the dark matter and baryon density fluctuations, respectively. In the large-scale limit the value of the transfer function reduces to logarithmic growth $f=d\ln{D}/d\ln{a} \sim 1$ with a value very close to unity at $z>2$ in the matter dominated era. On small scales ($k\gg k_F$) the shape of $T_\eta$ inherits the asymptotic behaviour of the baryon density and decays as $T_\eta \propto k^{-2}$.

\begin{figure}
\centering
\includegraphics[width=0.45\textwidth]{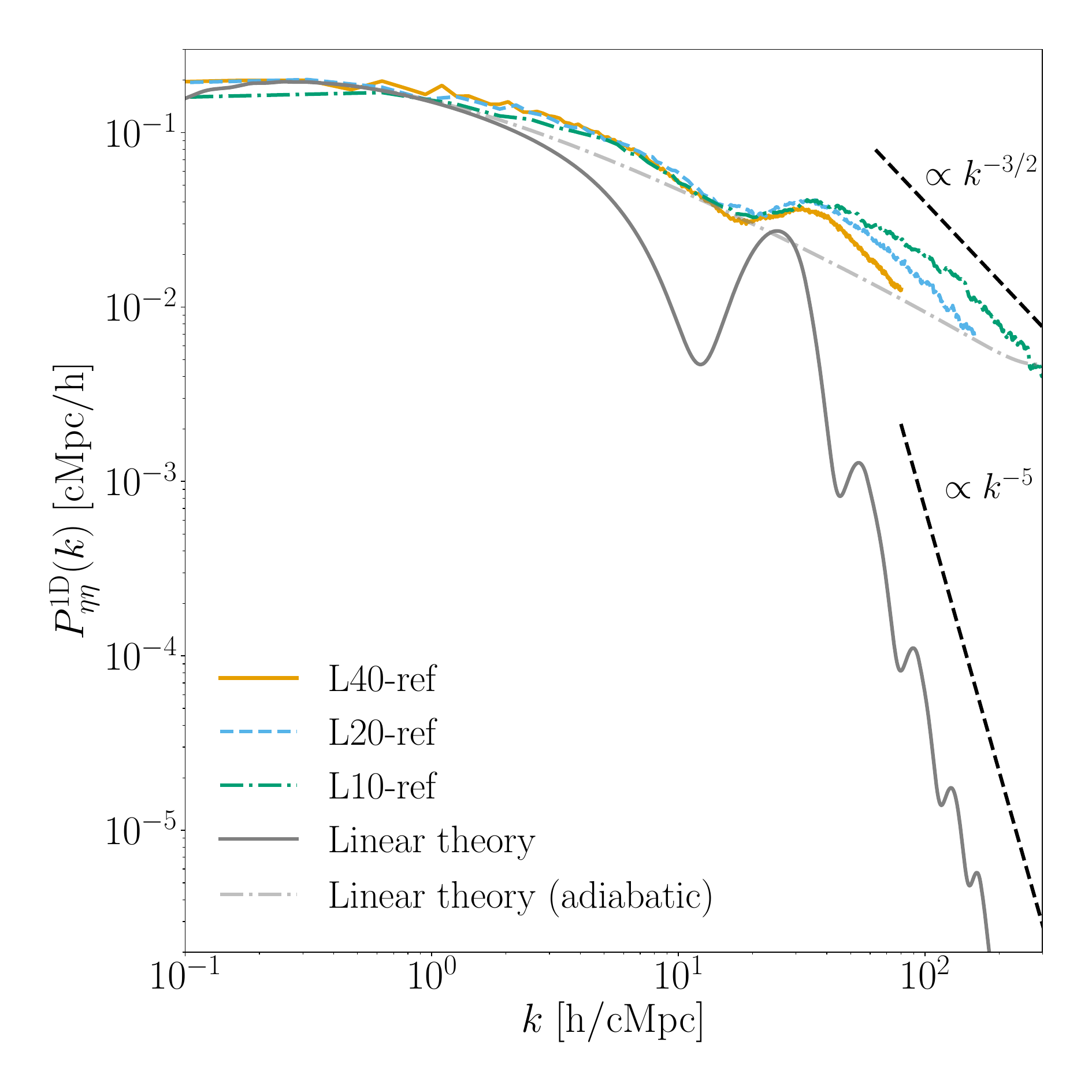}
\caption{Comparison of the velocity gradient power spectra from simulations with the linear theory prediction at $z=4.0$. The linear theory solution shows damped acoustic oscillations (solid gray) using the tabulated sound speed from the simulations, while in the absence of reionization (adiabatic cooling) the result of linear theory smoothly declines (dot-dashed gray). The position of the small-scale structure peak in the simulated $P_{\eta\eta}^{1D}$ coincides with the position of the first peak in the linear theory calculation. The asymptotic behaviour of the linear theory ($\propto k^{-5}$) and simulations ($\propto k^{-3/2}$) are indicated by black dashed lines. The asymptotic power-law is present in simulations across a wide range of numerical resolutions, from high resolution (L10-ref; dot-dashed green) to low resolution (L40-ref; solid orange). }
\label{fig:eta_linear_theory}
\end{figure}

Fig.~\ref{fig:eta_linear_theory} shows the resulting 1D power spectrum of the $\eta$ field from the linear theory calculation using the tabulated sound speeds as a function of time, as calculated for a gas parcel photo-heated and photo-ionized by the UV background used in the L20-ref simulation \citep{Puchwein19}. On large scales ($k<1\;h/\mathrm{cMpc}$) the agreement between simulations and linear theory is remarkably good, within 20\%. 

Despite the 1D power spectrum being a projected summary statistic, the acoustic oscillation pattern is still present at intermediate scales ($1 < k/[\mathrm{{{\it{h}}/cMpc}}] < 100$) in the $\eta$ field, with its first peak coinciding with the small-scale feature seen in $P_{\eta\eta}^{1D}$ measured in simulations. This result is distinct from a linear theory solution without reionization where adiabatic cooling persists to $z=4$ (dot-dashed gray line). The position and width of the peak are somewhat different in the simulations compared to the linear theory. This is likely the result of the simulations that capture the average effect of the linear theory solution where the local value of the sound speed varies with position (due to the inhomogeneity in the gas temperature and the electron fraction), while the linear solution in Fig.~\ref{fig:eta_linear_theory} reflects the result for the globally averaged sound speed \citep{Noaz2007}, evaluated at mean density. Additionally, much like baryon acoustic oscillations in the distribution of galaxies, nonlinear gravitational collapse is known to smear out the oscillations and reduce the amplitude in Fourier space \citep{Eisenstein2007,Smith2008}. 

On small scales ($k>100\;\mathrm{{\it{h}}/cMpc}$) the linear theory approaches an asymptotic behaviour that suppresses the power of $\eta$ as $k^{-5}$. In the power-law limit of $T_\eta \propto k^t$ and matter power spectrum $P_m \propto k^p$, the asymptotic behaviour of $P_{\eta\eta}^{1D} \propto k^{2+p+2t}$. In the high-k regime investigated here, the linear matter power is to a good approximation proportional to $k^{n_s-4} \sim k^{-3}$. For the case of no reionization and adiabatic cooling, the filtering scale is very large $k_F \sim 10^3\;\mathrm{{\it{h}}/cMpc}$, and therefore $T_\eta \propto k^0$, which results in the $P_{\eta\eta}^{1D} \propto k^{-1}$ asymptote (light gray). For the case where reionization is included in the sound speed, $k_F \sim 30\;\mathrm{{\it{h}}/cMpc}$, and therefore $T_\eta\propto k^{-2}$ at $k>100\;\mathrm{{\it{h}}/cMpc}$, which leads to $P_{\eta\eta}^{1D} \propto k^{-5}$.

Simulations on the other hand show an asymptote that sits somewhere in between these two limits at $P_{\eta\eta} \propto k^{-3/2}$. There could be several possible reasons for this difference. On one hand, the already mentioned non-linear collapse and its effect on the peculiar velocity field or averaging of the linear solution for inhomogenous sound speed could lead to different asymptotic behaviour in the high-k regime. Additionally, virial shocks in the simulation would likely produce a much broader feature in the Fourier space, and could contribute to the overall broader signal. And lastly, the rate of change of the sound speed could also contribute to the change in how fast the acoustic oscillations are damped. The asymptotic behaviour of $T_\eta \propto k^{-2}$ at $k>k_F$ is a particular solution of the baryon growth equation in the limit of constant or slowly varying sound speed \citep{Gnedin98}. The reionization process however can introduce sharp features in the sound speed variations. The traditional Jeans condition of $a\,H < k\,c_s$ leads to the transition from $\delta_b \propto k^0$ to $\propto k^{-2}$. However if $|dc_s/d\tau| \gg kc_s^2$ during reionization, where $\tau$ is conformal time and $c_s$ is the sound speed, then this leads to a transient feature that adds ringing to the solution with an envelope $\delta_b \propto k^{-n}$ and $n<2$. This is exactly what the filtering scale solution of \citet{Gnedin98} sought to capture. The linear theory solution for reionization does not change dramatically if relative baryon-dark matter velocities are included \citep{Tseliakhovich2010}, though its exact impact on simulations requires further study \citep{Cain2026}.

The linear theory solution supports the argument that the small-scale feature in $P_{\eta\eta}^{1D}$ in simulations is tracing the acoustic oscillations as induced by the time dependent sound speed that arises through the reionization process.

\section{Results of the simulations}
\label{sec:results}

Using the simulation setup described in the previous sections, the following sections systematically explore the evolution of the small-scale feature in the velocity and flux power spectra.

\begin{figure*}
\centering
    \includegraphics[width=0.45\textwidth]{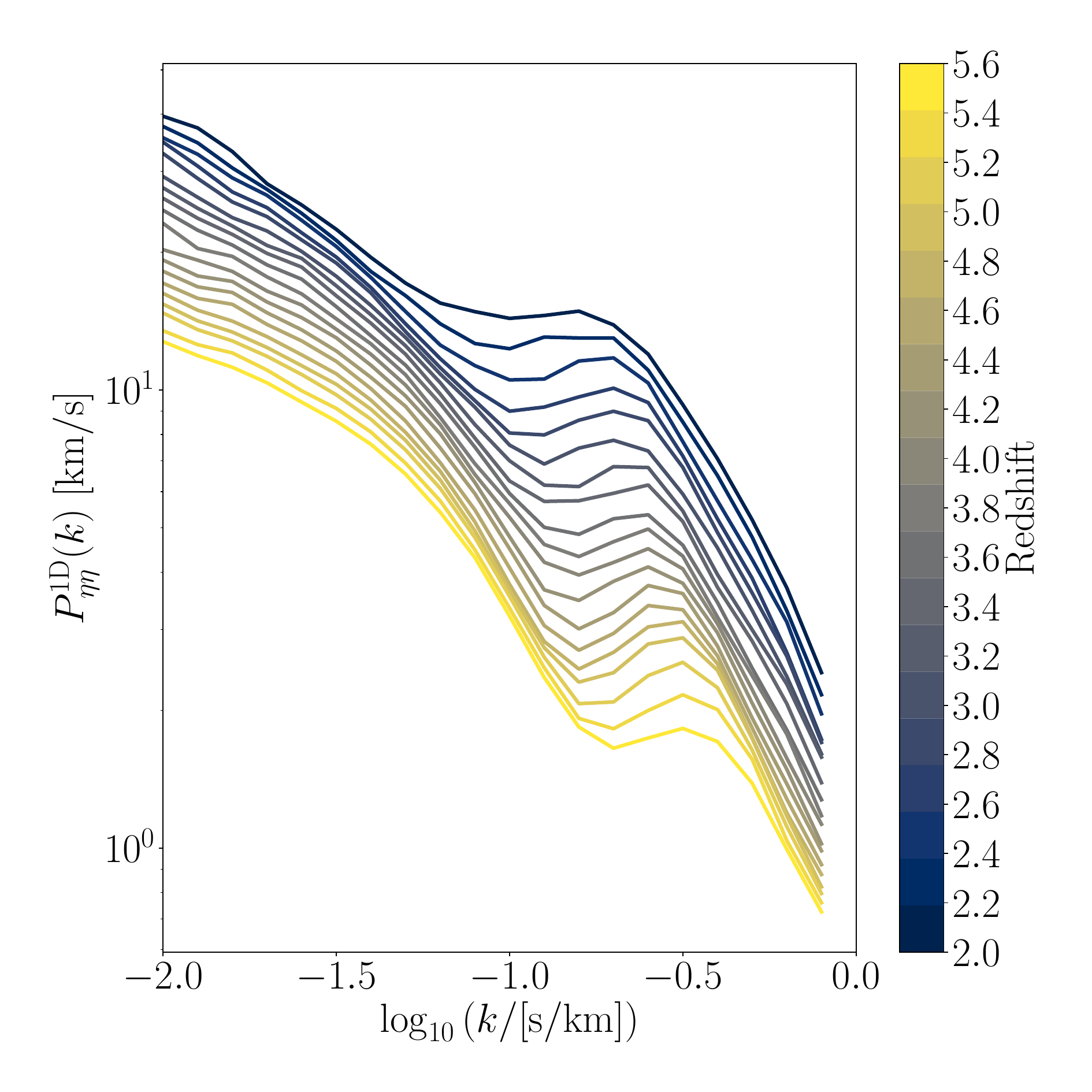}
    \includegraphics[width=0.45\textwidth]{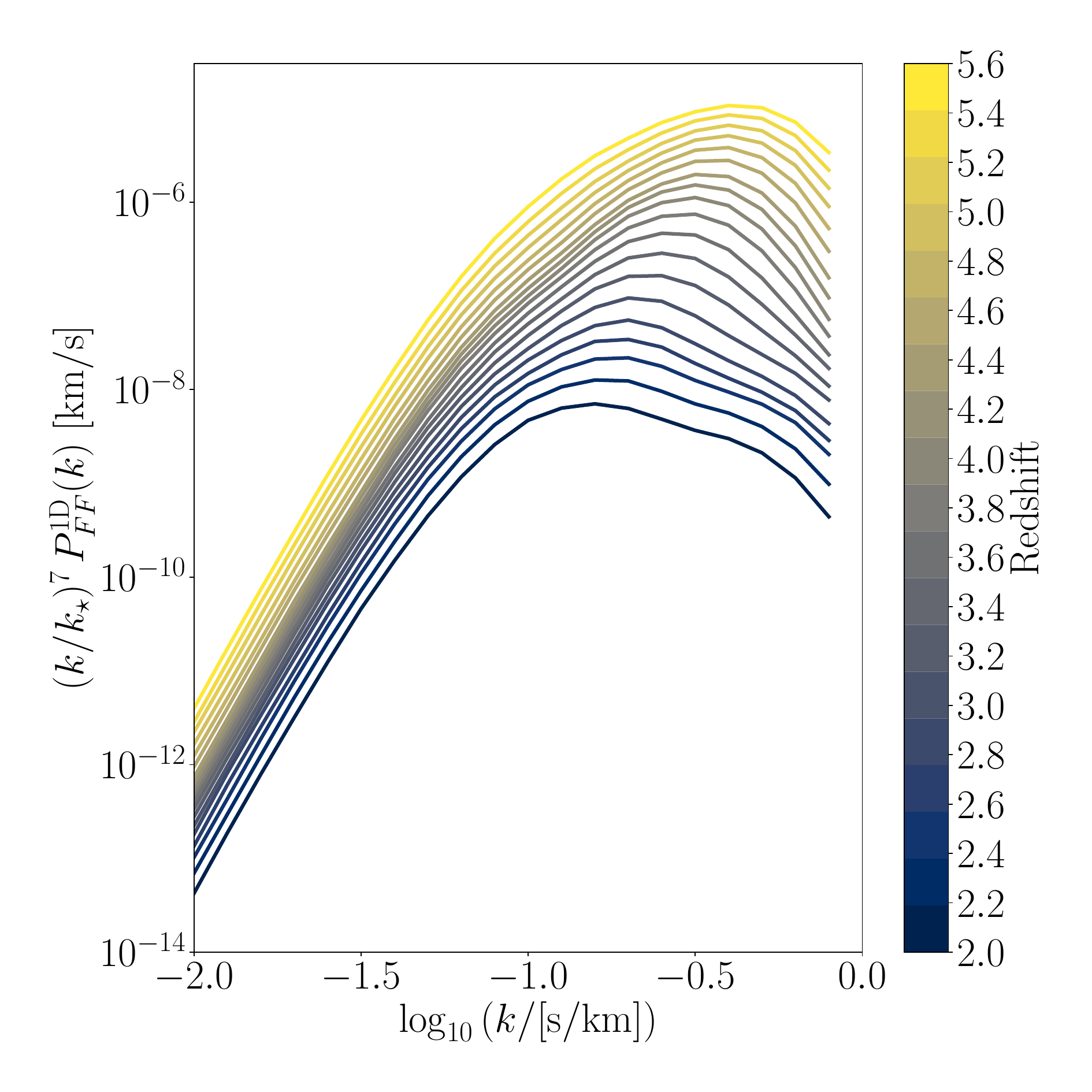}
\caption{The redshift evolution of the power spectra over the redshift range $z=2-5.6$ for the reference simulated model (L40-ref). {\it Left:} The power spectrum of velocity gradients along the line-of-sight shows similar evolution, with the position of the small-scale structure in the $\eta$ power spectrum systematically moving towards higher-k with increasing redshift. {\it Right:} The evolution of the flux power spectrum, highlighting the behaviour at small-scales. The small-scale feature in the flux power spectrum moves from larger to smaller scales with increasing redshift, tracking the evolution seen in the velocity space. The adopted value for the pivot scale is $k_\star=1\;\mathrm{s/km}$. The scaling was adopted to highlight the small-scale feature in the flux power spectrum. } \label{fig:redshift_evolution}
\end{figure*}

\subsection{Redshift evolution}
\label{sec:redshift_evolution}

\begin{figure}
\centering
\includegraphics[width=0.45\textwidth]{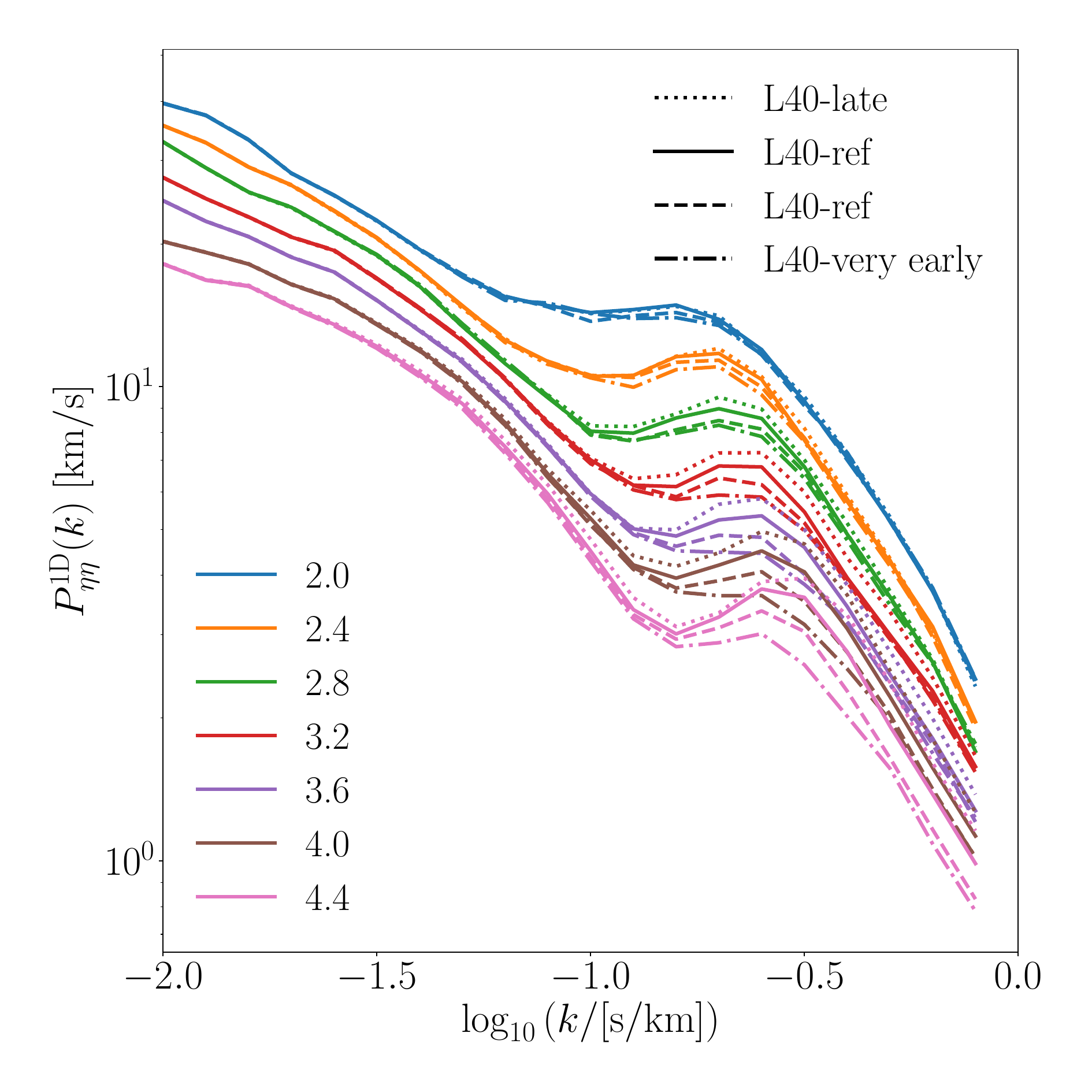}
\caption{The redshift evolution of the power spectrum of the gradient of the peculiar velocity field for different thermal history models with varying redshift of the end of the reionization: late ($z^{\rm end}_{\rm rei}\approx5.3$; dotted), mid ($z^{\rm end}_{\rm rei}\approx6.0$; solid), early ($z^{\rm end}_{\rm rei}\approx6.7$; dashed), and very early ($z^{\rm end}_{\rm rei}\approx7.4$; dot-dashed) reionization models of \citet{Puchwein2023}. The small-scale structure in the velocity space is more sensitive to the models of reionization at higher redshifts, affecting the shape, amplitude and position of the small-scale feature.}
\label{fig:eta_redshift_models}
\end{figure}

Expanding structures due to the hydrodynamic response of overpressurised gas should leave a redshift dependent imprint on the velocity and flux power spectra. Indeed, for an expanding sphere undergoing a hydrodynamic response, the expansion velocity is expected to decrease with time, following the decrease in the pressure gradients. The initial time evolution of the expansion will be sensitive to the details of the thermodynamics, such as the temperature and pressure of the gas. This early stage offers a unique window into the physics and timing of reionization process in the Universe. At a later time, after the hydrodynamic response has slowed down, the typical scale of the expanding structure continues to grow due to the expansion of the Universe. This later phase of the expansion can serve as a measure of the expansion rate that is independent of the baryonic acoustic oscillations.

Fig.~\ref{fig:redshift_evolution} shows the redshift evolution of the small-scale structure imprinted on the velocity power (left) and flux power spectra (right), respectively. The evolution is shown for the reference simulation (L40-ref) over a wide redshift range. While the small-scale feature is sharp and clearly distinguishable in the velocity power spectra, to highlight the same behaviour in the flux power spectra the figure shows $(k/k_\star)^7\,P_F(k)$ scaling, with $k_\star = 1\;h/\mathrm{cMpc}$. The choice of the pivot scale value of $k_\star$ and the power-law index is purely for visualization purposes. In both power spectra it is clear that the position of the peak features moves to larger scales with decreasing redshift, indicating a growth of the size of the responsible underlying structures.

Fig.~\ref{fig:eta_redshift_models} shows the redshift evolution of the velocity power spectra, but also this time for a variety of different thermal history models. The main difference between the models is the reionization history, with different models fully reionizing the Universe ($x_{\rm HII}=0.999$) at different times. The results nicely illustrate that at high redshifts the position and shape of the peak in the velocity power spectra depend significantly on the reionization history. This dependence on astrophysics is, however, erased at lower redshifts, long after the reionization is completed in any of the models.

\subsection{Astrophysical \& Cosmological information}
\label{sec:information}

To assess the sensitivity of this typical pressure scale to the underlying astrophysical and cosmological model, the redshift evolution from the previous section can be converted into how the typical scale of the effect evolves with redshift in both the velocity and flux power spectra. In this proof-of-concept study a simple set of analytical models is fit to the shape of both power spectra, in order to extract the exact position of the small-scale feature. While the shape of the empirical fitting function is split into a smooth and peak component, this does not follow a formal analytical derivation. The details of the two fitting functions are presented in Appendix~\ref{appendix:B}. Here we summarize some of the main features of the simulations that informed the functional form of the empirical relations.

\begin{figure*}
\centering
\includegraphics[width=0.45\textwidth]{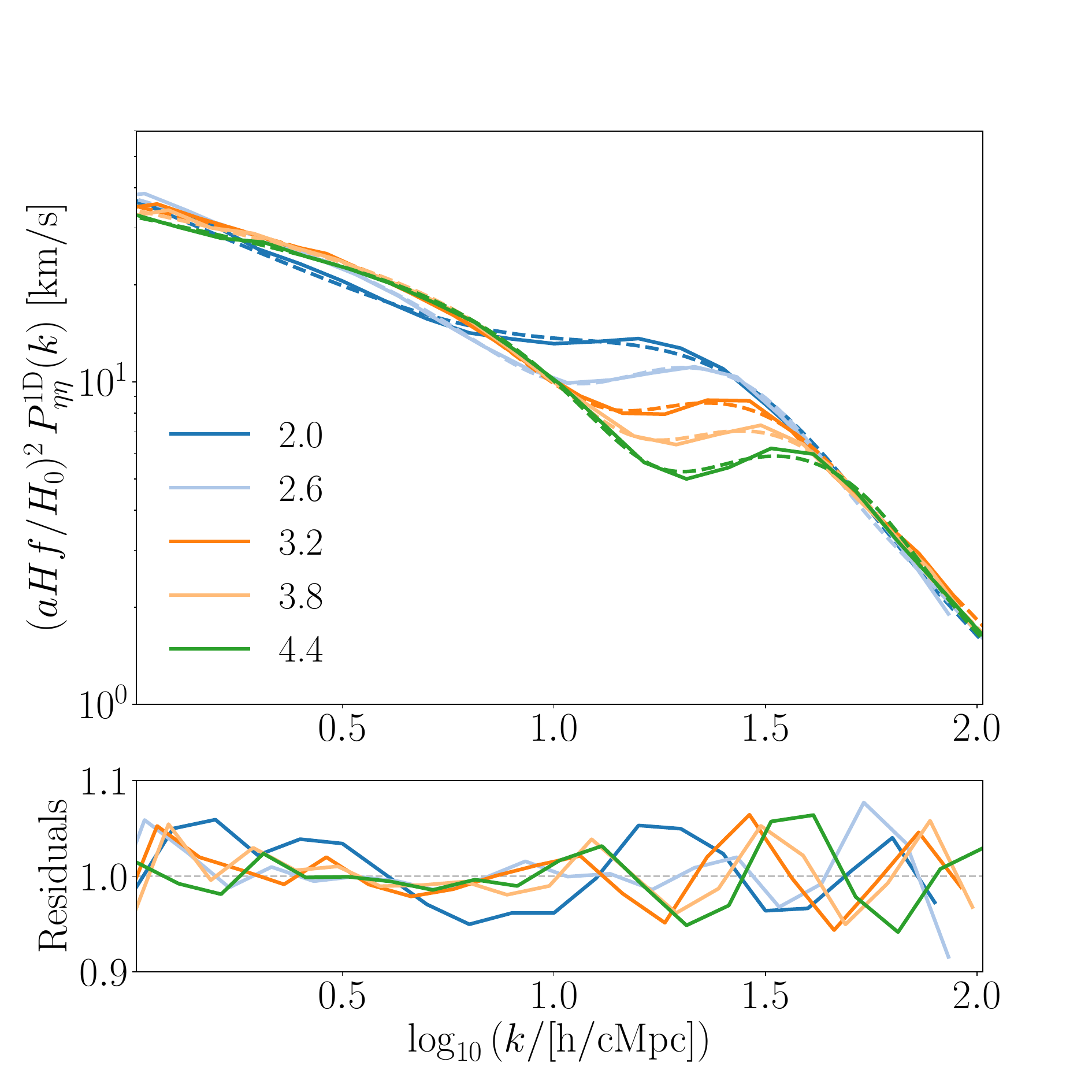}
\includegraphics[width=0.45\textwidth]{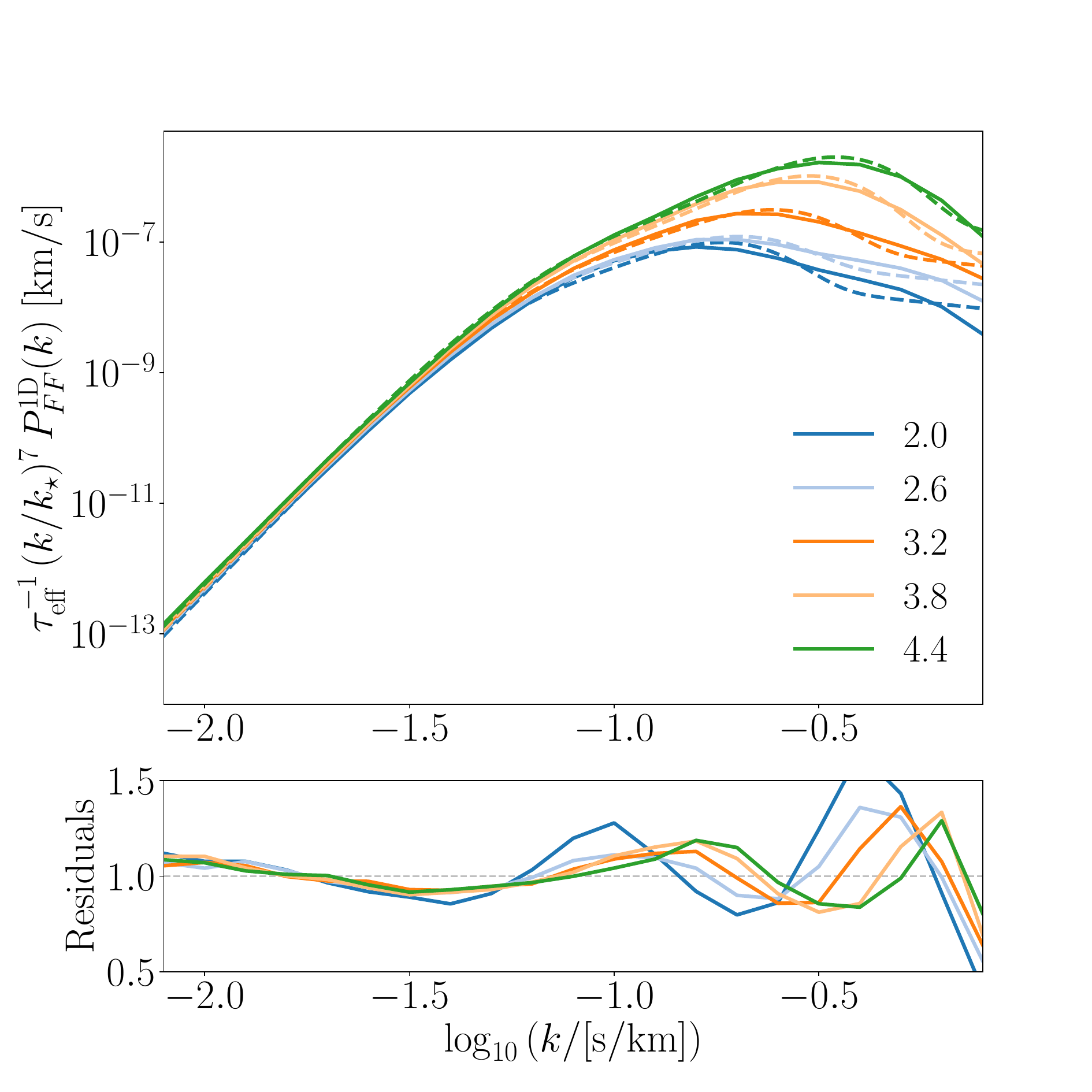}
\caption{Empirical fits to the velocity and flux power spectra (top panels) with residuals (bottom panels). In the top panels the solid lines show the results of the simulations and dashed line show the empirical fits. {\it Left:} The velocity gradient power spectrum on large scales traces the growth of density fluctuations in the linear regime through the continuity equation, such that $\eta \propto (aH\,f)$, and shows almost uniform power-law scaling at high-k as $P_{\eta\eta}\propto k^{-3/2}$ (see also Fig.~\ref{fig:eta_linear_theory}). {\it Right:} The flux power spectrum shows a transition in its steep small-scale suppression on scales corresponding to the feature in the velocity power. The effect is broader and less pronounced in the flux power spectra.}
\label{fig:fitting_models}
\end{figure*}

The velocity power spectra show a degree of self-similarity across the entire redshift range of $2<z<5$. By converting the velocity units of the wave vector $k$ to distance units, and multiplying the power spectra by $(f aH/H_0)^2$, the resulting velocity power spectra exhibit similar trends on scales both smaller and larger than the feature induced by the hydrodynamic response, as is shown in the top left panel of Fig.~\ref{fig:fitting_models}. This demonstrates that the amplitude of the velocity power spectra quite closely follows the expectation of linear theory, where the gradient of the peculiar velocity $\eta$ is proportional to the matter overdensity $\delta$, with the constant of proportionality given by the product of the logarithmic growth rate ($f={\rm d}\ln{D}/{\rm d}\ln{a}$) and the conformal Hubble expansion rate ($aH$). This constant of proportionality encodes, to a large degree, the redshfit evolution of the velocity power amplitude across a vast range of scales and cosmic times.

The main deviation from the near-universal redshift scaling is related to the redshift evolution of the small-scale structure associated with the small-scale hydrodynamic response. The shape of this feature connects the large- and small-scale behaviour of the velocity flux power spectra at $\sim$ few tens of $h/$cMpc. The comoving size of the feature also clearly changes as a function of redshift. The peak of the feature becomes broader as redshift decreases, and the prominence of the feature starts to disappear and move from a peak to a knee in the velocity power.

The second deviation, while smaller in scope, is that on large scales the velocity power does not follow the linear theory scaling with redshift exactly. This could be due to a couple of different reasons: the expected mode coupling due to the inherent non-linearity of the gravitational collapse, and due to limited box size of the simulations which at $40\;\mathrm{cMpc/h}$ are not sufficiently large to be expected to follow linear theory at the scales corresponding to the size of the simulation volume.

A rather remarkable result is also the behaviour of the velocity power spectrum in the high-k limit, where the power scales as a power-law, $P_{\eta\eta} \propto k^{-3/2}$, across the entire redshift range. This statement is related to the discussion in the literature on whether the ratio of baryon and dark matter density fluctuations approaches a fixed power-law in the large k regime. As argued in \citet{Gnedin98}, such asymptotic behaviour should emerge only for very special time evolution of the temperature with redshift ($T \propto (1+z)^\beta$). Nevertheless, this is seen here in a hydrodynamical simulation with much more complex gas temperature evolution.

The flux power spectra also show a certain degree of self-similarity, although much more approximately than in the case of the velocities. The rescaling in this case is only applied on the amplitude of the flux power spectrum, which was multiplied by $\tau_{\rm eff}^{-1}$. The resulting flux power spectrum shows roughly self-similar behaviour at large scales, with $\tau_{\rm eff}^{-1}\,P_F^{1D} \propto k^{-1/2}$ at $k<0.02\;\mathrm{s/km}$, with a much slower transition towards the small scale structure. A peak in $\propto k^7\,P_F^{1D}(k)$ appears at similar scales as the feature in $P_{\eta\eta}^{1D}$, as shown in the top right panel of Fig.~\ref{fig:fitting_models}.

The position of the small-scale feature is less well discerned in the flux power spectrum, as it has a much lower amplitude than in the velocity power spectrum. The feature is also much broader in wavenumber space in the flux power spectrum, but it has a similar trend of being more prominent at higher redshifts and disappearing at low redshifts. Unlike the velocity power spectra, the flux power spectra at different redshift exhibit more complex behaviour at small scales, beyond the position of the feature ($k>0.4\;\skm$). The goodness-of-fit to the flux power spectrum does not depend on the exact nature of the power-law scaling ($k^7$), and the choice was made for visualization purposes only.

The subsequent peak positions in the model are not directly given by the scale of the peak position in the model $k_p$, as such statement would be heavily dependent on the exact functional forms of the model. Instead, the peak positions in velocity and flux power spectra ($k_p^\eta$,$k_p^F$) are determined by finding the local small-scale maxima of $(aHf)^2\,P_{\eta\eta}^{1D}$ and $\tau_{\rm eff}^{-1}\,k^7\,P_F^{1D}$ models respectively. The uncertainty of the fitting parameters, as well as determining the position of the local maxima, has been propagated into the respective uncertainties for ($k_p^\eta$,$k_p^F$). 

\section{Pressure smoothing measurements}
\label{sec:measurements}

\begin{figure*}
\centering
\includegraphics[width=0.45\textwidth]{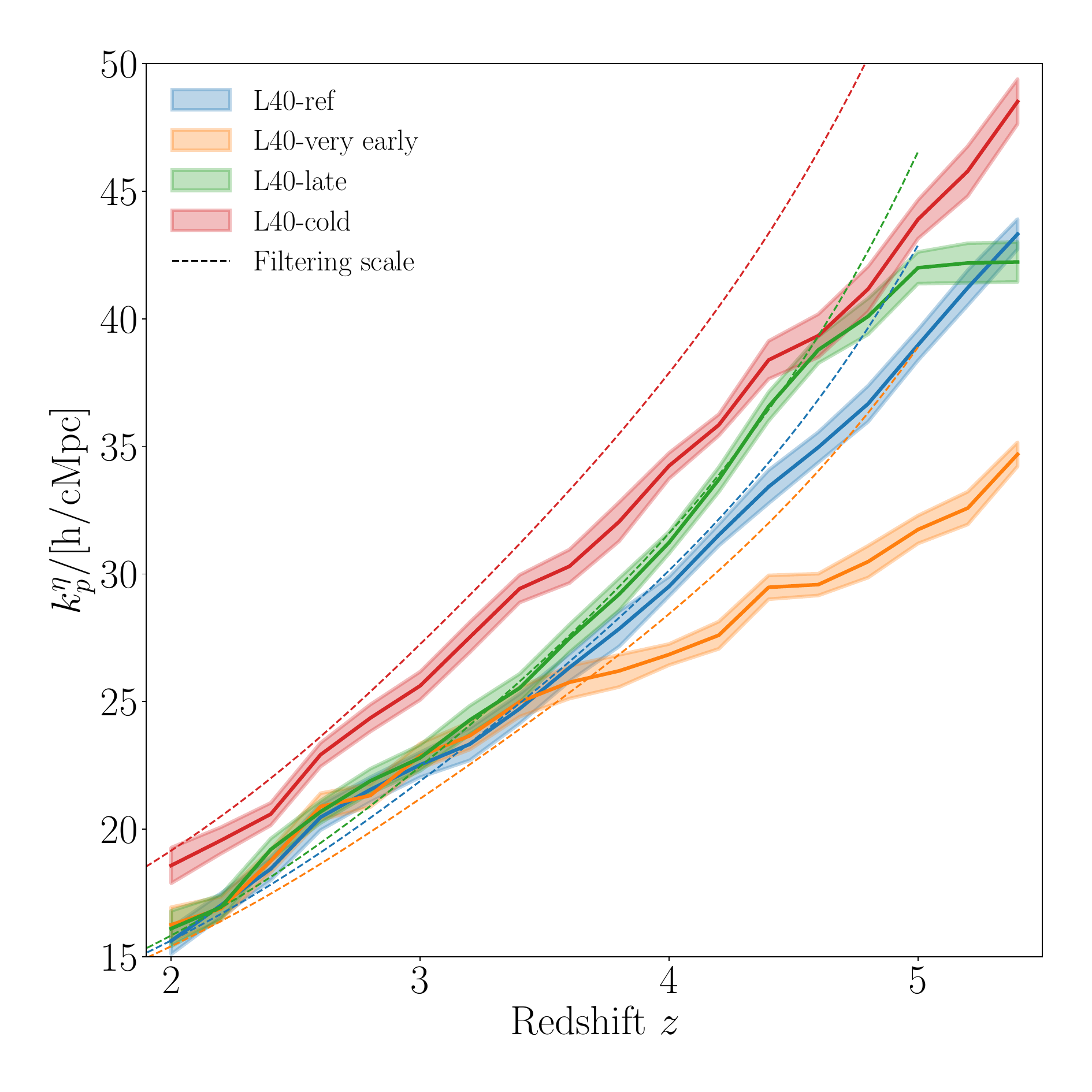}
\includegraphics[width=0.45\textwidth]{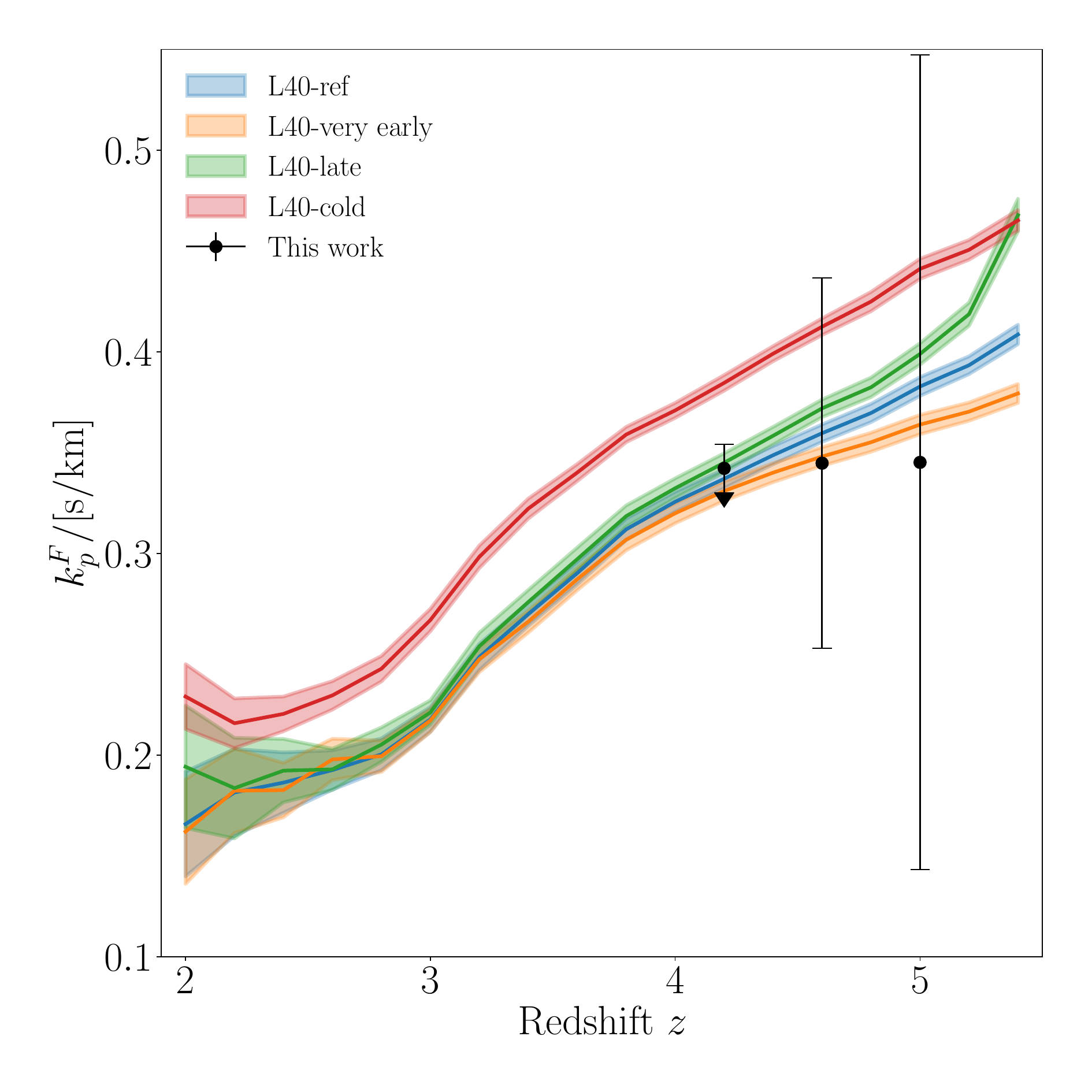}
\caption{The position of the power spectrum peak as a function of redshift. The coloured bands represent the $1\sigma$ uncertainty of the peak position estimation (solid line), propagating the uncertainty of the empirical fit parameters. {\it Left:} The position of the peak in the velocity gradient power spectrum for different simulated models \citep{Puchwein2023}: reference model with $z_{\rm rei}^{\rm end} \approx 6$ (L40-ref, blue); a colder thermal history model with the same end redshift of reionization (L40-cold, red); and two models with later (L40-late, green) and earlier (L40-very early, orange) end of reionization with $z_{\rm rei}^{\rm end} \approx 5.3$ and $7.4$, respectively. For each simulation the dashed lines also show the filtering scale \citep{Gnedin98}, evaluated at mean density for each model. {\it Right:} The position of the peak in the flux power spectrum ($\propto k^7\,P_F$) for the same simulated thermal histories is shown in the left panel. Also overplotted are measurements from applying the same technique to the observed flux power spectrum measurements of \citet{Boera2019}.
}
\label{fig:standard_ruler}
\end{figure*}

The resulting redshift evolution of the peak position in the velocity ($k_p^\eta$) and flux ($k_p^F$) power spectra are shown in Fig.~\ref{fig:standard_ruler}, with the peak position in Fourier space moving towards larger scales with decreasing redshift. Different colours represent results from simulations with different thermal histories. Both the velocity and flux peak positions diverge at high redshifts for models where reionization ends at different times; while the same models show remarkable agreement in the peak position at lower redshifts, long after reionization has ended. Similarly, models with the same reionization history but different gas temperature in the simulation, show only a difference in the value of the peak position, while the redshift dependence remains almost identical. For the model with lower gas temperature (red) compared to the reference model (blue), the value of the corresponding $k_p^i$, for $i\in{\eta,F}$, is larger, indicating that the size of the underlying expanding structure is smaller. This is expected if the expansion velocity is proportional to the sound speed in the intergalactic medium where $c_s \propto \sqrt{T}$. 

At low redshifts different models overlap and the redshift evolution of the peak predominantly traces the expansion history of the Universe. At high redshifts the dependence of the peak position on the thermal history is more pronounced, with models that have reionized late showing a faster expansion of the peak position from small to large scales than models that have reionized early. 

As a comparison, Fig.~\ref{fig:standard_ruler} also shows the redshift evolution as predicted by the filtering scale \citep{Gnedin98}. The filtering scale depends only on the expansion history of the Universe and the redshift evolution of the sound speed in the simulations, and has been evaluated at mean density. The peak position in the velocity power spectra shows generally good agreement with the predictions of the filtering scale, suggesting that the position of the peak in the velocity power spectra is tightly linked to the pressure smoothing scale, as was shown in Section~\ref{sec:linear_theory}. There are however, some differences, in particular towards high redshifts, where the slope of the redshift evolution in simulations changes compared to the predictions of \citep{Gnedin98}. This difference is more acute for models where reionization has ended early, and the gas has had longer to expand.

On the other hand, the redshift evolution of the peak position as estimated from the flux power spectra shows a slightly different redshift evolution. While above general remarks still hold, including the sensitivity to thermal history at high redshifts and the dependence on the gas temperature, the redshift evolution of all models at $z>3$ suggests a shallower power-law evolution, $k_p^F \propto (1+z)^q$, with $q<1$. The position of the peak in the velocity power spectra fits the evolution better with $q>1$\footnote{The sudden uptick/downtick of the model where reionization ends late (green; zr525) is at least partially due to attempting to estimate the hydrodynamic response before reionization has finished in that particular model. The exact position measured would also depend highly on the patchy nature of reioniozation that we otherwise neglect in this work.}. The position of the peak also becomes difficult to measure at lower redshifts in the flux power spectra, where the peak is much broader in Fourier space. As a result the size of the peak estimated from the flux power flattens for all models at $z<2.8$.

An important caveat of the estimates in this work is that the peak position is at relatively small scales (e.g. $k_{\rm max} = 0.8\;\mathrm{s/km}$ for the flux power spectrum). While accessible in simulations, such high wavenumbers are prohibitive to measure in real data, due to a variety of factors including the effects of instrumental noise, instrument resolution and contamination by clustered metal lines. Currently the best measurements in the high-k regime extend to $k_{\rm max} = 0.2\skm$ \citep{Boera2019}, that have been recently used to put tight constraints on e.g. different dark matter models or reionization histories \citep{Boera2019,Rogers2021,Villasenor2022b,Irsic2024,GarciaGallego2025b}. However, these high-k range of observations is still susceptible to residual systematics from the correlated metals, instrumental resolution and noise modelling \citep{Ma2026,Irsic2024,Boera2019} that will need to be assessed in more detail when the precision on the pressure smoothing scale measurement decreases with future data.

Applying the fitting techniques using the smooth/peak model for the flux power to the measurements of \citep{Boera2019} results in the first direct measurements of the pressure smoothing scale at high redshift. The results are shown as black points in Fig.~\ref{fig:standard_ruler}. The fit to the data also took into account the observed flux power spectrum uncertainties. Using the full observed covariance of \citet{Irsic2024} had little impact on the best-fit. The empirical fit was estimated with a gradient-based decent method, where the Jacobian was used as an estimate of the parameter covariance. The peak position was estimated as the local maximum in the empirical fit at small scales, folding the uncertainties on the fit parameters into the final summarized errors in Table.~\ref{tab:lambda_pressure}.

As the empirical fits were applied directly to the data, no additional resolution correction due to numerical convergence was applied. The current observational uncertainties on $\lambda_p = 2\pi/k_p^F$ (Table.~\ref{tab:lambda_pressure}) are dominated by the flux power uncertainty and the value of $k_{\rm max}$ in real data. Other important contributions are from the poor performance of the empirical fit on simulated flux power ($\leq50\%$) and the general broadness of the feature in the flux field. Further studies going beyond this proof-of-concept result will likely have to employ more strict and model independent methodology to determine the position of the small-scale peak in $k^7\,P_F(k)$, for example following methods developed for the Baryon Acoustic Oscillations \citep{Paranjape2026}.

\section{Discussion}
\label{sec:discussion}

\begin{figure}
\centering
\includegraphics[width=0.45\textwidth]{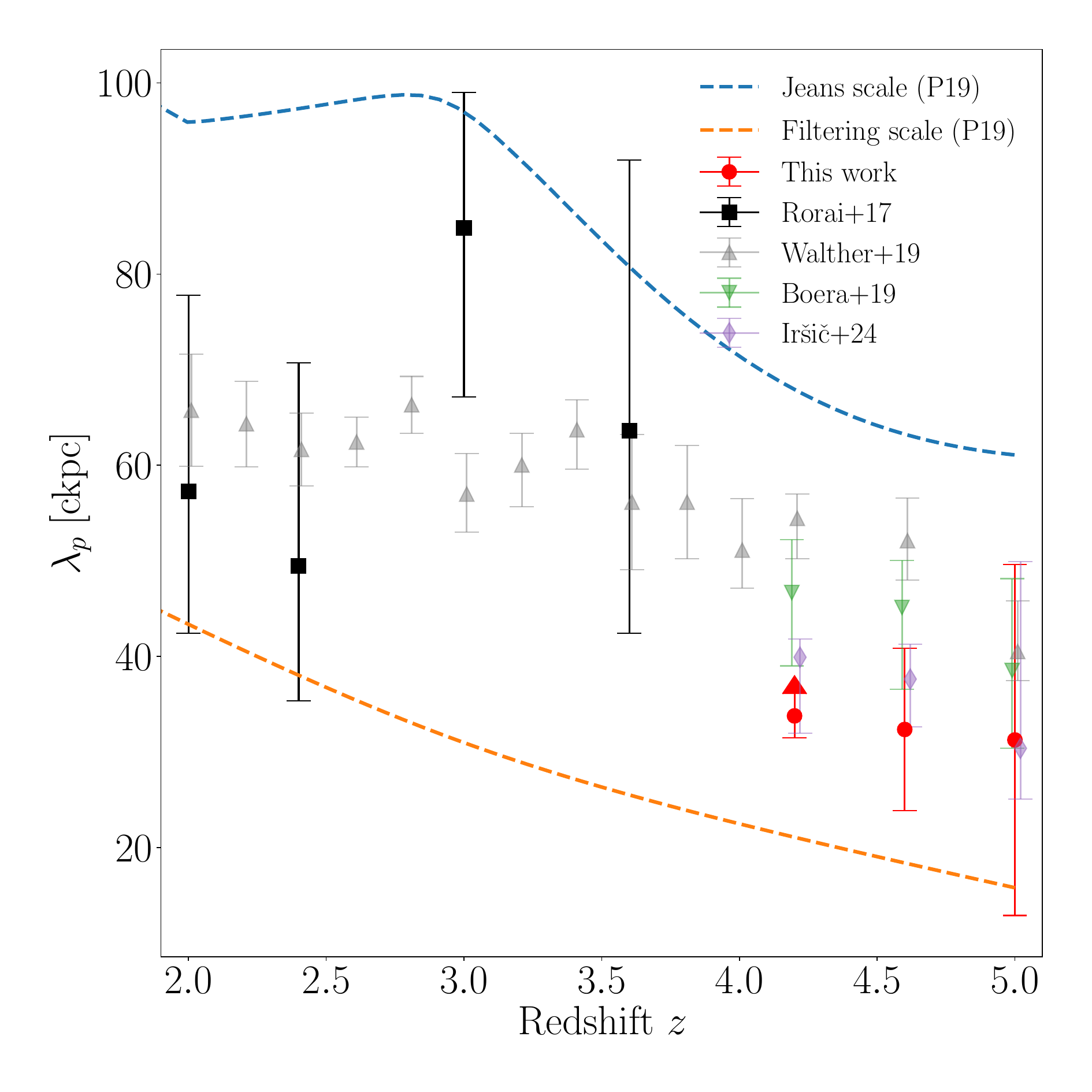}
\caption{The redshift evolution of the inferred pressure smoothing scale as measured through this work, using high redshift Lyman-$\alpha$ forest 1D flux power spectrum measurements of \citep{Boera2019} (red points). Previous direct measurements of the pressure smoothing scale have covered lower redshift \citep{Rorai2017} (black points). Additionally previous indirect measurements through P1D small-scale power suppression by \citep{Walther2018} (gray points); \citep{Boera2019} (green points); and \citep{Irsic2024} (purple points). Typically the pressure smoothing scale should be smaller than the Jeans scale due to memory of the thermal and ionization history \citep{Gnedin98}, calculated for the thermal evolution of our mid reionization model \citep{Puchwein2023} (blue-dashed). In reality the pressure smoothing scale should be closer to, and possibly larger than, the linear theory estimates \citep{Gnedin98} (orange-dashed).}
\label{fig:lambda_p}
\end{figure}

These results add to the limited sample of direct measurements of the pressure smoothing scale. Whereas previous measurements \citep{Rorai2017} used the phase correlation between nearby quasar sightlines, the results presented in this work access the information of the pressure smoothing through the imprint of the peculiar velocities the small-scale flux clustering. The result presented in this paper showcases for the first time such measurements at $z>4$. The results are summarized in Fig.~\ref{fig:lambda_p}. In contrast, most previous studies of the pressure smoothing effect in the IGM used indirect methods, relying on measuring the effect of cumulative heat injection and its effect on the flux power spectrum on much larger scales \citep{Walther2018,Boera2019,Onorbe2019,Garzilli2021}. At such larger scales, any method used needs to disentangle the different physical effects that produce a cutoff in the flux power (thermal broadening, Jeans pressure, warm dark matter free streaming). This is typically done by marginalizing over a relatively large range of different thermal histories and cosmological models. The method presented here bypasses this by focussing directly on the physical feature produced by the gas pressure. The provided measurements can in turn be used as a physically informed prior for the full shape 1D flux power analyses, resulting in tighter constraints on the derived parameters.

\begin{table}
    \centering
    \begin{tabular}{c|c}
        Redshift & $\lambda_p\;\mathrm{[ckpc]}$ \\\hline
        $z=4.2$ & $>33.78$ ($1\sigma$) \\
        $z=4.6$ & $32.36^{+8.35}_{-8.35}$ \\
        $z=5.0$ & $31.27^{+18.28}_{-18.28}$
    \end{tabular}
    \caption{Measured pressure smoothing scale using the small scale feature in the flux power spectra of \citep{Boera2019}. The origin of the small scale feature can be traced to the hydrodynamic response of the overpressurized gas after reionization, as imprinted on the peculiar velocity structure traced by the observed flux. }
    \label{tab:lambda_pressure}
\end{table}

Fig.~\ref{fig:lambda_p} compares the redshift evolution of the pressure smoothing scale measurements from a collection of these different sources. The direct measurements are in general agreement with each other, and the expected redshift evolution of the linear theory filtering scaling from \citep{Gnedin98}. A caveat of both the observational uncertainties and the methodology employed is that in its current state the direct measurements at $z>4$ are unable to differentiate between different thermal histories.

The indirect measurements \citep{Walther2018,Boera2019} are similarly in good agreement with the alternative methods that directly measure the pressure smoothing scale. In particular, at $z>4$ different methods that are based off of the same \citep{Boera2019} flux power spectrum measurements are all within $1\sigma$. This statement is even stronger for the indirect analysis of the flux power spectrum measurements with the same simulation setup \citep{Irsic2024}. Nevertheless, some tension with previous measurements is observed \citep{Walther2018}, in particular at $z=4.6$. These older measurements were reliant on \citep{Viel13wdm} HIRES/MIKE spectra that had access to lower observed $k_{\rm max}$, which might have contributed to difference compared to the newer analysis of \citep{Boera2019}.

\section{Summary and conclusions}
\label{sec:summary}

The thermal history of the Universe traces the last known phase transition in the Universe -- the process of reionization. Its impact on early galaxy formation and evolution and large-scale structure underpins the importance of large observational and theoretical programmes to understand how and when the Universe reionized. After the process of reionization is largely complete, the cooling of the intergalactic medium slowly erases the information about the thermal history, while the remaining signal is mostly locked in the integrated effect of the pressure smoothing. This effect captures the hydrodynamic response of the gas to the overpressurized system left in the wake of ionization fronts. 

This work develops intuitive toy and linear theory models to explain the effect of the hydrodynamic response left in the power spectrum of the gradient of the peculiar velocity field, which in turn affects the observed flux in distant quasar spectra. The signal is characterized and explored both across cosmic time and across our simulation suite designed to encapsule viable reionization history models. 

Analysing high-resolution hydrodynamical simulations, this work provides a connection between the effects of the hydrodynamic response and reionization as imprinted on the peculiar velocity field, and how this effect is propagated to the small-scale structure of the power spectrum of the observed transmission in the quasar flux. While the small-scale flux power spectrum is sensitive to a variety of subtle effects \citep{Cain2024,Cain2026}, a significant contribution is imprinted by the peculiar velocity structure that traces the predictions of linear theory \citep{Gnedin98}. A toy model where expanding spheres were injected into a smoothed velocity field of the simulation can explain the signal in both the velocity and flux power spectra.

The redshift evolution of the flux and velocity power spectra show cohesive change in the position of the small-scale feature that appears as a peak in $P_{\eta\eta}^{1D}$ and $k^7\,P_{F}^{1D}$ at scales of $0.2-0.8\;\skm$, with the value increasing with increasing redshift. Smooth empirical functions were used to fit the simulations around the peak to provide a robust estimate of the peak position in both flux and velocity space, with its redshift evolution tracing that of the filtering scale prediction \citep{Gnedin98} using the sound speed evolution of the simulations \citep{Puchwein2023}. Different simulated models closely agree at $z<3.5$ and trace the Hubble expansion rate, while they diverge at higher redshifts closer to the end redshift of reionization in the models.

The same technique to recover the peak position was applied to the observed flux power spectrum measurements of \citep{Boera2019}, providing first, direct measurements of the pressure smoothing scale at high redshifts ($z>4.2$). The measurements presented in this work broadly agree with previous direct detection measurements \citep{Rorai2017} as well as indirect measurements that rely on the overall shape of the flux power spectrum \citep{Walther2018,Boera2019}. While the current uncertainties make model inference incapable of discriminating between the models, future observations of the flux power spectra that can push to higher $k_{\rm max}$ (e.g. GHOSTLy \citep{Artola2024}, EQUALS \citep{Berg2025}) will offer a new window into direct measurements of the relics of reionization, and effects of the heat injection during the entire history of this last phase transition.

\section*{Acknowledgements}
VI acknowledges support by the Kavli Foundation. MV is supported by the INFN PD51 INDARK grant and by the INAF Theory Grant "Cosmological investigation of the cosmic web". Support by ERC Advanced Grant 320596 ‘The Emergence of Structure During the Epoch of Reionization’ is gratefully acknowledged. MGH has been supported by STFC consolidated grant ST/N000927/1 and ST/S000623/1. JSB is supported by STFC consolidated grants ST/T000171/1 and ST/X000982/1. LIG was supported as a summer intern at University of Hertfordshire. For the purpose of open access, the author has applied a Creative Commons Attribution (CC BY) licence to any Author Accepted Manuscript version arising from this submission.

The simulations used in this work were performed using the Joliot Curie supercomputer at the Tré Grand Centre de Calcul (TGCC) and the Cambridge Service for Data Driven Discovery (CSD3), part of which is operated by the University of Cambridge Research Computing on behalf of the STFC DiRAC HPC Facility (www.dirac.ac.uk). We acknowledge the Partnership for Advanced Computing in Europe (PRACE) for awarding us time on Joliot Curie in the 16th call. The DiRAC component of CSD3 was funded by BEIS capital funding via STFC capital grants ST/P002307/1 and ST/R002452/1 and STFC operations grant ST/R00689X/1. This work also used the DiRAC@Durham facility managed by the Institute for Computational Cosmology on behalf of the STFC DiRAC HPC Facility. The equipment was funded by BEIS capital funding via STFC capital grants ST/P002293/1 and ST/R002371/1, Durham University and STFC operations grant ST/R000832/1. DiRAC is part of the National e-Infrastructure.

\section*{Data Availability}
The data and analysis code used in this work are available from the authors on request.  Further guidance for accessing  the publicly available Sherwood-Relics simulation data can be found on the project website:  \url{https://www.nottingham.ac.uk/astronomy/sherwood-relics/}

%%%%%%%%%%%%%%%%%%%%%%%%%%%%%%%%%%%%%%%%%%%%%%%%%%



\bibliographystyle{mnras}
\bibliography{references} % if your bibtex file is called example.bib

@ARTICLE{rorai13,
       author = {{Rorai}, Alberto and {Hennawi}, Joseph F. and {White}, Martin},
        title = "{A New Method to Directly Measure the Jeans Scale of the Intergalactic Medium Using Close Quasar Pairs}",
      journal = {\apj},
         year = 2013,
        month = oct,
       volume = {775},
       number = {2},
          eid = {81},
        pages = {81},
          doi = {10.1088/0004-637X/775/2/81},
archivePrefix = {arXiv},
       eprint = {1305.0210},
 primaryClass = {astro-ph.CO},
       adsurl = {https://ui.adsabs.harvard.edu/abs/2013ApJ...775...81R}
}

@ARTICLE{kulkarni15,
       author = {{Kulkarni}, Girish and {Hennawi}, Joseph F. and {O{\~n}orbe}, Jose and {Rorai}, Alberto and {Springel}, Volker},
        title = "{Characterizing the Pressure Smoothing Scale of the Intergalactic Medium}",
      journal = {\apj},
         year = 2015,
        month = oct,
       volume = {812},
       number = {1},
          eid = {30},
        pages = {30},
          doi = {10.1088/0004-637X/812/1/30},
archivePrefix = {arXiv},
       eprint = {1504.00366},
 primaryClass = {astro-ph.CO},
       adsurl = {https://ui.adsabs.harvard.edu/abs/2015ApJ...812...30K}
}

@ARTICLE{Boera2019,
       author = {{Boera}, Elisa and {Becker}, George D. and {Bolton}, James S. and {Nasir}, Fahad},
        title = "{Revealing Reionization with the Thermal History of the Intergalactic Medium: New Constraints from the Ly{\ensuremath{\alpha}} Flux Power Spectrum}",
      journal = {\apj},
         year = 2019,
        month = feb,
       volume = {872},
       number = {1},
          eid = {101},
        pages = {101},
          doi = {10.3847/1538-4357/aafee4},
archivePrefix = {arXiv},
       eprint = {1809.06980},
 primaryClass = {astro-ph.CO},
       adsurl = {https://ui.adsabs.harvard.edu/abs/2019ApJ...872..101B}
}

@ARTICLE{Irsic2024,
       author = {{Ir{\v{s}}i{\v{c}}}, Vid and {Viel}, Matteo and {Haehnelt}, Martin G. and {Bolton}, James S. and {Molaro}, Margherita and {Puchwein}, Ewald and {Boera}, Elisa and {Becker}, George D. and {Gaikwad}, Prakash and {Keating}, Laura C. and {Kulkarni}, Girish},
        title = "{Unveiling dark matter free streaming at the smallest scales with the high redshift Lyman-alpha forest}",
      journal = {\prd},
         year = 2024,
        month = feb,
       volume = {109},
       number = {4},
          eid = {043511},
        pages = {043511},
          doi = {10.1103/PhysRevD.109.043511},
archivePrefix = {arXiv},
       eprint = {2309.04533},
 primaryClass = {astro-ph.CO},
       adsurl = {https://ui.adsabs.harvard.edu/abs/2024PhRvD.109d3511I}
}

@ARTICLE{Doughty2023,
       author = {{Doughty}, Caitlin C. and {Hennawi}, Joseph F. and {Davies}, Frederick B. and {Luki{\'c}}, Zarija and {O{\~n}orbe}, Jose},
        title = "{Convergence of small scale Ly{\ensuremath{\alpha}} structure at high-z under different reionization scenarios}",
      journal = {\mnras},
         year = 2023,
        month = aug,
          doi = {10.1093/mnras/stad2549},
archivePrefix = {arXiv},
       eprint = {2305.16200},
 primaryClass = {astro-ph.CO},
       adsurl = {https://ui.adsabs.harvard.edu/abs/2023MNRAS.tmp.2467D}
}

@ARTICLE{palanque20,
       author = {{Palanque-Delabrouille}, Nathalie and {Y{\`e}che}, Christophe and {Sch{\"o}neberg}, Nils and {Lesgourgues}, Julien and {Walther}, Michael and {Chabanier}, Sol{\`e}ne and {Armengaud}, Eric},
        title = "{Hints, neutrino bounds, and WDM constraints from SDSS DR14 Lyman-{\ensuremath{\alpha}} and Planck full-survey data}",
      journal = {\jcap},
         year = 2020,
        month = apr,
       volume = {2020},
       number = {4},
          eid = {038},
        pages = {038},
          doi = {10.1088/1475-7516/2020/04/038},
archivePrefix = {arXiv},
       eprint = {1911.09073},
 primaryClass = {astro-ph.CO},
       adsurl = {https://ui.adsabs.harvard.edu/abs/2020JCAP...04..038P}
}

@ARTICLE{springel05,
       author = {{Springel}, Volker},
        title = "{The cosmological simulation code GADGET-2}",
      journal = {\mnras},
         year = 2005,
        month = dec,
       volume = {364},
       number = {4},
        pages = {1105-1134},
          doi = {10.1111/j.1365-2966.2005.09655.x},
archivePrefix = {arXiv},
       eprint = {astro-ph/0505010},
 primaryClass = {astro-ph},
       adsurl = {https://ui.adsabs.harvard.edu/abs/2005MNRAS.364.1105S}
}

@ARTICLE{Planck2020,
       author = {{Planck Collaboration} and {Aghanim}, N. and {Akrami}, Y. and {Ashdown}, M. and {Aumont}, J. and {Baccigalupi}, C. and {Ballardini}, M. and {Banday}, A.~J. and {Barreiro}, R.~B. and {Bartolo}, N. and {Basak}, S. and {Battye}, R. and {Benabed}, K. and {Bernard}, J. -P. and {Bersanelli}, M. and {Bielewicz}, P. and {Bock}, J.~J. and {Bond}, J.~R. and {Borrill}, J. and {Bouchet}, F.~R. and {Boulanger}, F. and {Bucher}, M. and {Burigana}, C. and {Butler}, R.~C. and {Calabrese}, E. and {Cardoso}, J. -F. and {Carron}, J. and {Challinor}, A. and {Chiang}, H.~C. and {Chluba}, J. and {Colombo}, L.~P.~L. and {Combet}, C. and {Contreras}, D. and {Crill}, B.~P. and {Cuttaia}, F. and {de Bernardis}, P. and {de Zotti}, G. and {Delabrouille}, J. and {Delouis}, J. -M. and {Di Valentino}, E. and {Diego}, J.~M. and {Dor{\'e}}, O. and {Douspis}, M. and {Ducout}, A. and {Dupac}, X. and {Dusini}, S. and {Efstathiou}, G. and {Elsner}, F. and {En{\ss}lin}, T.~A. and {Eriksen}, H.~K. and {Fantaye}, Y. and {Farhang}, M. and {Fergusson}, J. and {Fernandez-Cobos}, R. and {Finelli}, F. and {Forastieri}, F. and {Frailis}, M. and {Fraisse}, A.~A. and {Franceschi}, E. and {Frolov}, A. and {Galeotta}, S. and {Galli}, S. and {Ganga}, K. and {G{\'e}nova-Santos}, R.~T. and {Gerbino}, M. and {Ghosh}, T. and {Gonz{\'a}lez-Nuevo}, J. and {G{\'o}rski}, K.~M. and {Gratton}, S. and {Gruppuso}, A. and {Gudmundsson}, J.~E. and {Hamann}, J. and {Handley}, W. and {Hansen}, F.~K. and {Herranz}, D. and {Hildebrandt}, S.~R. and {Hivon}, E. and {Huang}, Z. and {Jaffe}, A.~H. and {Jones}, W.~C. and {Karakci}, A. and {Keih{\"a}nen}, E. and {Keskitalo}, R. and {Kiiveri}, K. and {Kim}, J. and {Kisner}, T.~S. and {Knox}, L. and {Krachmalnicoff}, N. and {Kunz}, M. and {Kurki-Suonio}, H. and {Lagache}, G. and {Lamarre}, J. -M. and {Lasenby}, A. and {Lattanzi}, M. and {Lawrence}, C.~R. and {Le Jeune}, M. and {Lemos}, P. and {Lesgourgues}, J. and {Levrier}, F. and {Lewis}, A. and {Liguori}, M. and {Lilje}, P.~B. and {Lilley}, M. and {Lindholm}, V. and {L{\'o}pez-Caniego}, M. and {Lubin}, P.~M. and {Ma}, Y. -Z. and {Mac{\'\i}as-P{\'e}rez}, J.~F. and {Maggio}, G. and {Maino}, D. and {Mandolesi}, N. and {Mangilli}, A. and {Marcos-Caballero}, A. and {Maris}, M. and {Martin}, P.~G. and {Martinelli}, M. and {Mart{\'\i}nez-Gonz{\'a}lez}, E. and {Matarrese}, S. and {Mauri}, N. and {McEwen}, J.~D. and {Meinhold}, P.~R. and {Melchiorri}, A. and {Mennella}, A. and {Migliaccio}, M. and {Millea}, M. and {Mitra}, S. and {Miville-Desch{\^e}nes}, M. -A. and {Molinari}, D. and {Montier}, L. and {Morgante}, G. and {Moss}, A. and {Natoli}, P. and {N{\o}rgaard-Nielsen}, H.~U. and {Pagano}, L. and {Paoletti}, D. and {Partridge}, B. and {Patanchon}, G. and {Peiris}, H.~V. and {Perrotta}, F. and {Pettorino}, V. and {Piacentini}, F. and {Polastri}, L. and {Polenta}, G. and {Puget}, J. -L. and {Rachen}, J.~P. and {Reinecke}, M. and {Remazeilles}, M. and {Renzi}, A. and {Rocha}, G. and {Rosset}, C. and {Roudier}, G. and {Rubi{\~n}o-Mart{\'\i}n}, J.~A. and {Ruiz-Granados}, B. and {Salvati}, L. and {Sandri}, M. and {Savelainen}, M. and {Scott}, D. and {Shellard}, E.~P.~S. and {Sirignano}, C. and {Sirri}, G. and {Spencer}, L.~D. and {Sunyaev}, R. and {Suur-Uski}, A. -S. and {Tauber}, J.~A. and {Tavagnacco}, D. and {Tenti}, M. and {Toffolatti}, L. and {Tomasi}, M. and {Trombetti}, T. and {Valenziano}, L. and {Valiviita}, J. and {Van Tent}, B. and {Vibert}, L. and {Vielva}, P. and {Villa}, F. and {Vittorio}, N. and {Wandelt}, B.~D. and {Wehus}, I.~K. and {White}, M. and {White}, S.~D.~M. and {Zacchei}, A. and {Zonca}, A.},
        title = "{Planck 2018 results. VI. Cosmological parameters}",
      journal = {\aap},
         year = 2020,
        month = sep,
       volume = {641},
          eid = {A6},
        pages = {A6},
          doi = {10.1051/0004-6361/201833910},
archivePrefix = {arXiv},
       eprint = {1807.06209},
 primaryClass = {astro-ph.CO},
       adsurl = {https://ui.adsabs.harvard.edu/abs/2020A&A...641A...6P}
}

@ARTICLE{Puchwein2023,
       author = {{Puchwein}, Ewald and {Bolton}, James S. and {Keating}, Laura C. and {Molaro}, Margherita and {Gaikwad}, Prakash and {Kulkarni}, Girish and {Haehnelt}, Martin G. and {Ir{\v{s}}i{\v{c}}}, Vid and {{\v{S}}oltinsk{\'y}}, Tom{\'a}{\v{s}} and {Viel}, Matteo and {Aubert}, Dominique and {Becker}, George D. and {Meiksin}, Avery},
        title = "{The Sherwood-Relics simulations: overview and impact of patchy reionization and pressure smoothing on the intergalactic medium}",
      journal = {\mnras},
         year = 2023,
        month = mar,
       volume = {519},
       number = {4},
        pages = {6162-6183},
          doi = {10.1093/mnras/stac3761},
archivePrefix = {arXiv},
       eprint = {2207.13098},
 primaryClass = {astro-ph.CO},
       adsurl = {https://ui.adsabs.harvard.edu/abs/2023MNRAS.519.6162P}
}

@ARTICLE{Rogers2021,
       author = {{Rogers}, Keir K. and {Peiris}, Hiranya V.},
        title = "{Strong Bound on Canonical Ultralight Axion Dark Matter from the Lyman-Alpha Forest}",
      journal = {\prl},
         year = 2021,
        month = feb,
       volume = {126},
       number = {7},
          eid = {071302},
        pages = {071302},
          doi = {10.1103/PhysRevLett.126.071302},
archivePrefix = {arXiv},
       eprint = {2007.12705},
 primaryClass = {astro-ph.CO},
       adsurl = {https://ui.adsabs.harvard.edu/abs/2021PhRvL.126g1302R}
}

@ARTICLE{Garzilli2021,
       author = {{Garzilli}, Antonella and {Magalich}, Andrii and {Ruchayskiy}, Oleg and {Boyarsky}, Alexey},
        title = "{How to constrain warm dark matter with the Lyman-{\ensuremath{\alpha}} forest}",
      journal = {\mnras},
         year = 2021,
        month = apr,
       volume = {502},
       number = {2},
        pages = {2356-2363},
          doi = {10.1093/mnras/stab192},
       adsurl = {https://ui.adsabs.harvard.edu/abs/2021MNRAS.502.2356G}
}

@ARTICLE{Villasenor2022b,
       author = {{Villasenor}, Bruno and {Robertson}, Brant and {Madau}, Piero and {Schneider}, Evan},
        title = "{New Constraints on Warm Dark Matter from the Lyman-$\alpha$ Forest Power Spectrum}",
      journal = {arXiv e-prints},
         year = 2022,
        month = sep,
          eid = {arXiv:2209.14220},
        pages = {arXiv:2209.14220},
          doi = {10.48550/arXiv.2209.14220},
archivePrefix = {arXiv},
       eprint = {2209.14220},
 primaryClass = {astro-ph.CO},
       adsurl = {https://ui.adsabs.harvard.edu/abs/2022arXiv220914220V}
}

@article{Gnedin98,
      author         = "Gnedin, Nickolay Y. and Hui, Lam",
      title          = "{Probing the universe with the Lyman alpha forest: 1.
                        Hydrodynamics of the low density IGM}",
      journal        = "Mon. Not. Roy. Astron. Soc.",
      volume         = "296",
      year           = "1998",
      pages          = "44-55",
      doi            = "10.1046/j.1365-8711.1998.01249.x",
      eprint         = "astro-ph/9706219",
      archivePrefix  = "arXiv",
      primaryClass   = "astro-ph",
      reportNumber   = "FERMILAB-PUB-97-212-A",
      SLACcitation   = "%%CITATION = ASTRO-PH/9706219;%%"
}

@ARTICLE{Hui97,
   author = {{Hui}, L. and {Gnedin}, N.~Y.},
    title = "{Equation of state of the photoionized intergalactic medium}",
  journal = {\mnras},
   eprint = {astro-ph/9612232},
     year = 1997,
    month = nov,
   volume = 292,
    pages = {27},
   adsurl = {http://adsabs.harvard.edu/abs/1997MNRAS.292...27H}
}

@ARTICLE{Nasir16,
       author = {{Nasir}, Fahad and {Bolton}, James S. and {Becker}, George D.},
        title = "{Inferring the IGM thermal history during reionization with the Lyman {\ensuremath{\alpha}} forest power spectrum at redshift z ≃ 5}",
      journal = {\mnras},
         year = 2016,
        month = dec,
       volume = {463},
       number = {3},
        pages = {2335-2347},
          doi = {10.1093/mnras/stw2147},
archivePrefix = {arXiv},
       eprint = {1605.04155},
 primaryClass = {astro-ph.CO},
       adsurl = {https://ui.adsabs.harvard.edu/abs/2016MNRAS.463.2335N}
}

@ARTICLE{kulkarni19,
       author = {{Kulkarni}, Girish and {Keating}, Laura C. and {Haehnelt}, Martin G. and {Bosman}, Sarah E.~I. and {Puchwein}, Ewald and {Chardin}, Jonathan and {Aubert}, Dominique},
        title = "{Large Ly {\ensuremath{\alpha}} opacity fluctuations and low CMB {\ensuremath{\tau}} in models of late reionization with large islands of neutral hydrogen extending to z < 5.5}",
      journal = {\mnras},
         year = 2019,
        month = may,
       volume = {485},
       number = {1},
        pages = {L24-L28},
          doi = {10.1093/mnrasl/slz025},
archivePrefix = {arXiv},
       eprint = {1809.06374},
 primaryClass = {astro-ph.CO},
       adsurl = {https://ui.adsabs.harvard.edu/abs/2019MNRAS.485L..24K}
}

@ARTICLE{Gaikwad21,
       author = {{Gaikwad}, Prakash and {Srianand}, Raghunathan and {Haehnelt}, Martin G. and {Choudhury}, Tirthankar Roy},
        title = "{A consistent and robust measurement of the thermal state of the IGM at 2 {\ensuremath{\leq}} z {\ensuremath{\leq}} 4 from a large sample of Ly {\ensuremath{\alpha}} forest spectra: evidence for late and rapid He II reionization}",
      journal = {\mnras},
         year = 2021,
        month = sep,
       volume = {506},
       number = {3},
        pages = {4389-4412},
          doi = {10.1093/mnras/stab2017},
archivePrefix = {arXiv},
       eprint = {2009.00016},
 primaryClass = {astro-ph.CO},
       adsurl = {https://ui.adsabs.harvard.edu/abs/2021MNRAS.506.4389G}
}

@ARTICLE{Bosman2021,
       author = {{Bosman}, Sarah E.~I. and {Davies}, Frederick B. and {Becker}, George D. and {Keating}, Laura C. and {Davies}, Rebecca L. and {Zhu}, Yongda and {Eilers}, Anna-Christina and {D'Odorico}, Valentina and {Bian}, Fuyan and {Bischetti}, Manuela and {Cristiani}, Stefano V. and {Fan}, Xiaohui and {Farina}, Emanuele P. and {Haehnelt}, Martin G. and {Hennawi}, Joseph F. and {Kulkarni}, Girish and {Mesinger}, Andrei and {Meyer}, Romain A. and {Onoue}, Masafusa and {Pallottini}, Andrea and {Qin}, Yuxiang and {Ryan-Weber}, Emma and {Schindler}, Jan-Torge and {Walter}, Fabian and {Wang}, Feige and {Yang}, Jinyi},
        title = "{Hydrogen reionisation ends by $z=5.3$: Lyman-$\alpha$ optical depth measured by the XQR-30 sample}",
      journal = {arXiv e-prints},
         year = 2021,
        month = aug,
          eid = {arXiv:2108.03699},
        pages = {arXiv:2108.03699},
          doi = {10.48550/arXiv.2108.03699},
archivePrefix = {arXiv},
       eprint = {2108.03699},
 primaryClass = {astro-ph.CO},
       adsurl = {https://ui.adsabs.harvard.edu/abs/2021arXiv210803699B}
}

@ARTICLE{Onorbe2019,
       author = {{O{\~n}orbe}, Jose and {Davies}, F.~B. and {Luki{\'c}} and {}, Z. and {Hennawi}, J.~F. and {Sorini}, D.},
        title = "{Inhomogeneous reionization models in cosmological hydrodynamical simulations}",
      journal = {\mnras},
         year = 2019,
        month = jul,
       volume = {486},
       number = {3},
        pages = {4075-4097},
          doi = {10.1093/mnras/stz984},
archivePrefix = {arXiv},
       eprint = {1810.11683},
 primaryClass = {astro-ph.CO},
       adsurl = {https://ui.adsabs.harvard.edu/abs/2019MNRAS.486.4075O}
}

@ARTICLE{Keating2019,
       author = {{Keating}, Laura C. and {Weinberger}, Lewis H. and {Kulkarni}, Girish and {Haehnelt}, Martin G. and {Chardin}, Jonathan and {Aubert}, Dominique},
        title = "{Long troughs in the Lyman-{\ensuremath{\alpha}} forest below redshift 6 due to islands of neutral hydrogen}",
      journal = {\mnras},
         year = 2020,
        month = jan,
       volume = {491},
       number = {2},
        pages = {1736-1745},
          doi = {10.1093/mnras/stz3083},
archivePrefix = {arXiv},
       eprint = {1905.12640},
 primaryClass = {astro-ph.CO},
       adsurl = {https://ui.adsabs.harvard.edu/abs/2020MNRAS.491.1736K}
}

@ARTICLE{Gaikwad20,
       author = {{Gaikwad}, Prakash and {Rauch}, Michael and {Haehnelt}, Martin G. and {Puchwein}, Ewald and {Bolton}, James S. and {Keating}, Laura C. and {Kulkarni}, Girish and {Ir{\v{s}}i{\v{c}}}, Vid and {Ba{\~n}ados}, Eduardo and {Becker}, George D. and {Boera}, Elisa and {Zahedy}, Fakhri S. and {Chen}, Hsiao-Wen and {Carswell}, Robert F. and {Chardin}, Jonathan and {Rorai}, Alberto},
        title = "{Probing the thermal state of the intergalactic medium at z > 5 with the transmission spikes in high-resolution Ly {\ensuremath{\alpha}} forest spectra}",
      journal = {\mnras},
         year = 2020,
        month = jun,
       volume = {494},
       number = {4},
        pages = {5091-5109},
          doi = {10.1093/mnras/staa907},
archivePrefix = {arXiv},
       eprint = {2001.10018},
 primaryClass = {astro-ph.CO},
       adsurl = {https://ui.adsabs.harvard.edu/abs/2020MNRAS.494.5091G}
}

@ARTICLE{Becker13,
       author = {{Becker}, George D. and {Hewett}, Paul C. and {Worseck}, G{\'a}bor and {Prochaska}, J. Xavier},
        title = "{A refined measurement of the mean transmitted flux in the Ly{\ensuremath{\alpha}} forest over 2 < z < 5 using composite quasar spectra}",
      journal = {\mnras},
         year = 2013,
        month = apr,
       volume = {430},
       number = {3},
        pages = {2067-2081},
          doi = {10.1093/mnras/stt031},
archivePrefix = {arXiv},
       eprint = {1208.2584},
 primaryClass = {astro-ph.CO},
       adsurl = {https://ui.adsabs.harvard.edu/abs/2013MNRAS.430.2067B}
}

@ARTICLE{Walther2018,
       author = {{Walther}, Michael and {Hennawi}, Joseph F. and {Hiss}, Hector and {O{\~n}orbe}, Jose and {Lee}, Khee-Gan and {Rorai}, Alberto and {O'Meara}, John},
        title = "{A New Precision Measurement of the Small-scale Line-of-sight Power Spectrum of the Ly{\ensuremath{\alpha}} Forest}",
      journal = {\apj},
         year = 2018,
        month = jan,
       volume = {852},
       number = {1},
          eid = {22},
        pages = {22},
          doi = {10.3847/1538-4357/aa9c81},
archivePrefix = {arXiv},
       eprint = {1709.07354},
 primaryClass = {astro-ph.CO},
       adsurl = {https://ui.adsabs.harvard.edu/abs/2018ApJ...852...22W}
}

@ARTICLE{Wilson2022,
       author = {{Wilson}, Bayu and {Ir{\v{s}}i{\v{c}}}, Vid and {McQuinn}, Matthew},
        title = "{A measurement of the Ly {\ensuremath{\beta}} forest power spectrum and its cross with the Ly {\ensuremath{\alpha}} forest in X-Shooter XQ-100}",
      journal = {\mnras},
         year = 2022,
        month = jan,
       volume = {509},
       number = {2},
        pages = {2423-2442},
          doi = {10.1093/mnras/stab3017},
archivePrefix = {arXiv},
       eprint = {2106.04837},
 primaryClass = {astro-ph.CO},
       adsurl = {https://ui.adsabs.harvard.edu/abs/2022MNRAS.509.2423W}
}

@ARTICLE{Molaro2023,
       author = {{Molaro}, Margherita and {Ir{\v{s}}i{\v{c}}}, Vid and {Bolton}, James S. and {Lieu}, Maggie and {Keating}, Laura C. and {Puchwein}, Ewald and {Haehnelt}, Martin G. and {Viel}, Matteo},
        title = "{Possible evidence for a large-scale enhancement in the Lyman-{\ensuremath{\alpha}} forest power spectrum at redshift z {\ensuremath{\geq}} 4}",
      journal = {\mnras},
         year = 2023,
        month = may,
       volume = {521},
       number = {1},
        pages = {1489-1501},
          doi = {10.1093/mnras/stad598},
archivePrefix = {arXiv},
       eprint = {2303.05167},
 primaryClass = {astro-ph.CO},
       adsurl = {https://ui.adsabs.harvard.edu/abs/2023MNRAS.521.1489M}
}

@ARTICLE{Molaro2022,
       author = {{Molaro}, Margherita and {Ir{\v{s}}i{\v{c}}}, Vid and {Bolton}, James S. and {Keating}, Laura C. and {Puchwein}, Ewald and {Gaikwad}, Prakash and {Haehnelt}, Martin G. and {Kulkarni}, Girish and {Viel}, Matteo},
        title = "{The effect of inhomogeneous reionization on the Lyman {\ensuremath{\alpha}} forest power spectrum at redshift z > 4: implications for thermal parameter recovery}",
      journal = {\mnras},
         year = 2022,
        month = feb,
       volume = {509},
       number = {4},
        pages = {6119-6137},
          doi = {10.1093/mnras/stab3416},
archivePrefix = {arXiv},
       eprint = {2109.06897},
 primaryClass = {astro-ph.CO},
       adsurl = {https://ui.adsabs.harvard.edu/abs/2022MNRAS.509.6119M}
}

@ARTICLE{Bolton05,
       author = {{Bolton}, James S. and {Haehnelt}, Martin G. and {Viel}, Matteo and {Springel}, Volker},
        title = "{The Lyman {\ensuremath{\alpha}} forest opacity and the metagalactic hydrogen ionization rate at z\raisebox{-0.5ex}\textasciitilde 2-4}",
      journal = {\mnras},
         year = 2005,
        month = mar,
       volume = {357},
       number = {4},
        pages = {1178-1188},
          doi = {10.1111/j.1365-2966.2005.08704.x},
archivePrefix = {arXiv},
       eprint = {astro-ph/0411072},
 primaryClass = {astro-ph},
       adsurl = {https://ui.adsabs.harvard.edu/abs/2005MNRAS.357.1178B}
}

@ARTICLE{Miralda96,
   author = {{Miralda-Escud{\'e}}, J. and {Cen}, R. and {Ostriker}, J.~P. and 
	{Rauch}, M.},
    title = "{The Ly alpha Forest from Gravitational Collapse in the Cold Dark Matter + Lambda Model}",
  journal = {\apj},
   eprint = {astro-ph/9511013},
     year = 1996,
    month = nov,
   volume = 471,
    pages = {582},
      doi = {10.1086/177992},
   adsurl = {http://adsabs.harvard.edu/abs/1996ApJ...471..582M}
}

@ARTICLE{Peeples10a,
   author = {{Peeples}, M.~S. and {Weinberg}, D.~H. and {Dav{\'e}}, R. and 
	{Fardal}, M.~A. and {Katz}, N.},
    title = "{Pressure support versus thermal broadening in the Lyman {$\alpha$} forest - I. Effects of the equation of state on longitudinal structure}",
  journal = {\mnras},
archivePrefix = "arXiv",
   eprint = {0910.0256},
 primaryClass = "astro-ph.CO",
     year = 2010,
    month = may,
   volume = 404,
    pages = {1281-1294},
      doi = {10.1111/j.1365-2966.2010.16383.x},
   adsurl = {http://adsabs.harvard.edu/abs/2010MNRAS.404.1281P}
}

@ARTICLE{Peeples10b,
   author = {{Peeples}, M.~S. and {Weinberg}, D.~H. and {Dav{\'e}}, R. and 
	{Fardal}, M.~A. and {Katz}, N.},
    title = "{Pressure support versus thermal broadening in the Lyman {$\alpha$} forest - II. Effects of the equation of state on transverse structure}",
  journal = {\mnras},
archivePrefix = "arXiv",
   eprint = {0910.0250},
 primaryClass = "astro-ph.CO",
     year = 2010,
    month = may,
   volume = 404,
    pages = {1295-1305},
      doi = {10.1111/j.1365-2966.2010.16384.x},
   adsurl = {http://adsabs.harvard.edu/abs/2010MNRAS.404.1295P}
}

@ARTICLE{Upton15,
   author = {{Upton Sanderbeck}, P.~R. and {D'Aloisio}, A. and {McQuinn}, M.~J.
	},
    title = "{Models of the thermal evolution of the intergalactic medium after reionization}",
  journal = {\mnras},
archivePrefix = "arXiv",
   eprint = {1511.05992},
     year = 2016,
    month = aug,
   volume = 460,
    pages = {1885-1897},
      doi = {10.1093/mnras/stw1117},
   adsurl = {http://adsabs.harvard.edu/abs/2016MNRAS.460.1885U}
}

@ARTICLE{Puchwein15,
   author = {{Puchwein}, E. and {Bolton}, J.~S. and {Haehnelt}, M.~G. and 
	{Madau}, P. and {Becker}, G.~D. and {Haardt}, F.},
    title = "{The photoheating of the intergalactic medium in synthesis models of the UV background}",
  journal = {\mnras},
archivePrefix = "arXiv",
   eprint = {1410.1531},
     year = 2015,
    month = jul,
   volume = 450,
    pages = {4081-4097},
      doi = {10.1093/mnras/stv773},
   adsurl = {http://adsabs.harvard.edu/abs/2015MNRAS.450.4081P}
}

@ARTICLE{Theuns98,
   author = {{Theuns}, T. and {Leonard}, A. and {Efstathiou}, G. and {Pearce}, F.~R. and 
	{Thomas}, P.~A.},
    title = "{P\^{}3M-SPH simulations of the Lyalpha forest}",
  journal = {\mnras},
   eprint = {astro-ph/9805119},
     year = 1998,
    month = dec,
   volume = 301,
    pages = {478-502},
      doi = {10.1046/j.1365-8711.1998.02040.x},
   adsurl = {http://adsabs.harvard.edu/abs/1998MNRAS.301..478T}
}

@ARTICLE{Viel04,
   author = {{Viel}, M. and {Haehnelt}, M.~G. and {Springel}, V.},
    title = "{Inferring the dark matter power spectrum from the Lyman {$\alpha$} forest in high-resolution QSO absorption spectra}",
  journal = {\mnras},
   eprint = {astro-ph/0404600},
     year = 2004,
    month = nov,
   volume = 354,
    pages = {684-694},
      doi = {10.1111/j.1365-2966.2004.08224.x},
   adsurl = {http://adsabs.harvard.edu/abs/2004MNRAS.354..684V}
}

@ARTICLE{Puchwein19,
       author = {{Puchwein}, Ewald and {Haardt}, Francesco and {Haehnelt}, Martin G. and {Madau}, Piero},
        title = "{Consistent modelling of the meta-galactic UV background and the thermal/ionization history of the intergalactic medium}",
      journal = {\mnras},
         year = 2019,
        month = may,
       volume = {485},
       number = {1},
        pages = {47-68},
          doi = {10.1093/mnras/stz222},
archivePrefix = {arXiv},
       eprint = {1801.04931},
 primaryClass = {astro-ph.GA},
       adsurl = {https://ui.adsabs.harvard.edu/abs/2019MNRAS.485...47P}
}

@ARTICLE{planck18,
       author = {{Planck Collaboration} and {Aghanim}, N. and {Akrami}, Y. and {Ashdown}, M. and {Aumont}, J. and {Baccigalupi}, C. and {Ballardini}, M. and {Banday}, A.~J. and {Barreiro}, R.~B. and {Bartolo}, N. and {Basak}, S. and {Battye}, R. and {Benabed}, K. and {Bernard}, J. -P. and {Bersanelli}, M. and {Bielewicz}, P. and {Bock}, J.~J. and {Bond}, J.~R. and {Borrill}, J. and {Bouchet}, F.~R. and {Boulanger}, F. and {Bucher}, M. and {Burigana}, C. and {Butler}, R.~C. and {Calabrese}, E. and {Cardoso}, J. -F. and {Carron}, J. and {Challinor}, A. and {Chiang}, H.~C. and {Chluba}, J. and {Colombo}, L.~P.~L. and {Combet}, C. and {Contreras}, D. and {Crill}, B.~P. and {Cuttaia}, F. and {de Bernardis}, P. and {de Zotti}, G. and {Delabrouille}, J. and {Delouis}, J. -M. and {Di Valentino}, E. and {Diego}, J.~M. and {Dor{\'e}}, O. and {Douspis}, M. and {Ducout}, A. and {Dupac}, X. and {Dusini}, S. and {Efstathiou}, G. and {Elsner}, F. and {En{\ss}lin}, T.~A. and {Eriksen}, H.~K. and {Fantaye}, Y. and {Farhang}, M. and {Fergusson}, J. and {Fernandez-Cobos}, R. and {Finelli}, F. and {Forastieri}, F. and {Frailis}, M. and {Fraisse}, A.~A. and {Franceschi}, E. and {Frolov}, A. and {Galeotta}, S. and {Galli}, S. and {Ganga}, K. and {G{\'e}nova-Santos}, R.~T. and {Gerbino}, M. and {Ghosh}, T. and {Gonz{\'a}lez-Nuevo}, J. and {G{\'o}rski}, K.~M. and {Gratton}, S. and {Gruppuso}, A. and {Gudmundsson}, J.~E. and {Hamann}, J. and {Handley}, W. and {Hansen}, F.~K. and {Herranz}, D. and {Hildebrandt}, S.~R. and {Hivon}, E. and {Huang}, Z. and {Jaffe}, A.~H. and {Jones}, W.~C. and {Karakci}, A. and {Keih{\"a}nen}, E. and {Keskitalo}, R. and {Kiiveri}, K. and {Kim}, J. and {Kisner}, T.~S. and {Knox}, L. and {Krachmalnicoff}, N. and {Kunz}, M. and {Kurki-Suonio}, H. and {Lagache}, G. and {Lamarre}, J. -M. and {Lasenby}, A. and {Lattanzi}, M. and {Lawrence}, C.~R. and {Le Jeune}, M. and {Lemos}, P. and {Lesgourgues}, J. and {Levrier}, F. and {Lewis}, A. and {Liguori}, M. and {Lilje}, P.~B. and {Lilley}, M. and {Lindholm}, V. and {L{\'o}pez-Caniego}, M. and {Lubin}, P.~M. and {Ma}, Y. -Z. and {Mac{\'\i}as-P{\'e}rez}, J.~F. and {Maggio}, G. and {Maino}, D. and {Mandolesi}, N. and {Mangilli}, A. and {Marcos-Caballero}, A. and {Maris}, M. and {Martin}, P.~G. and {Martinelli}, M. and {Mart{\'\i}nez-Gonz{\'a}lez}, E. and {Matarrese}, S. and {Mauri}, N. and {McEwen}, J.~D. and {Meinhold}, P.~R. and {Melchiorri}, A. and {Mennella}, A. and {Migliaccio}, M. and {Millea}, M. and {Mitra}, S. and {Miville-Desch{\^e}nes}, M. -A. and {Molinari}, D. and {Montier}, L. and {Morgante}, G. and {Moss}, A. and {Natoli}, P. and {N{\o}rgaard-Nielsen}, H.~U. and {Pagano}, L. and {Paoletti}, D. and {Partridge}, B. and {Patanchon}, G. and {Peiris}, H.~V. and {Perrotta}, F. and {Pettorino}, V. and {Piacentini}, F. and {Polastri}, L. and {Polenta}, G. and {Puget}, J. -L. and {Rachen}, J.~P. and {Reinecke}, M. and {Remazeilles}, M. and {Renzi}, A. and {Rocha}, G. and {Rosset}, C. and {Roudier}, G. and {Rubi{\~n}o-Mart{\'\i}n}, J.~A. and {Ruiz-Granados}, B. and {Salvati}, L. and {Sandri}, M. and {Savelainen}, M. and {Scott}, D. and {Shellard}, E.~P.~S. and {Sirignano}, C. and {Sirri}, G. and {Spencer}, L.~D. and {Sunyaev}, R. and {Suur-Uski}, A. -S. and {Tauber}, J.~A. and {Tavagnacco}, D. and {Tenti}, M. and {Toffolatti}, L. and {Tomasi}, M. and {Trombetti}, T. and {Valenziano}, L. and {Valiviita}, J. and {Van Tent}, B. and {Vibert}, L. and {Vielva}, P. and {Villa}, F. and {Vittorio}, N. and {Wandelt}, B.~D. and {Wehus}, I.~K. and {White}, M. and {White}, S.~D.~M. and {Zacchei}, A. and {Zonca}, A.},
        title = "{Planck 2018 results. VI. Cosmological parameters}",
      journal = {\aap},
         year = 2020,
        month = sep,
       volume = {641},
          eid = {A6},
        pages = {A6},
          doi = {10.1051/0004-6361/201833910},
archivePrefix = {arXiv},
       eprint = {1807.06209},
 primaryClass = {astro-ph.CO},
       adsurl = {https://ui.adsabs.harvard.edu/abs/2020A&A...641A...6P}
}

@ARTICLE{lukic15,
       author = {{Luki{\'c}}, Zarija and {Stark}, Casey W. and {Nugent}, Peter and {White}, Martin and {Meiksin}, Avery A. and {Almgren}, Ann},
        title = "{The Lyman {\ensuremath{\alpha}} forest in optically thin hydrodynamical simulations}",
      journal = {\mnras},
         year = 2015,
        month = feb,
       volume = {446},
       number = {4},
        pages = {3697-3724},
          doi = {10.1093/mnras/stu2377},
archivePrefix = {arXiv},
       eprint = {1406.6361},
 primaryClass = {astro-ph.CO},
       adsurl = {https://ui.adsabs.harvard.edu/abs/2015MNRAS.446.3697L}
}

@ARTICLE{bolton17,
       author = {{Bolton}, James S. and {Puchwein}, Ewald and {Sijacki}, Debora and {Haehnelt}, Martin G. and {Kim}, Tae-Sun and {Meiksin}, Avery and {Regan}, John A. and {Viel}, Matteo},
        title = "{The Sherwood simulation suite: overview and data comparisons with the Lyman {\ensuremath{\alpha}} forest at redshifts 2 {\ensuremath{\leq}} z {\ensuremath{\leq}} 5}",
      journal = {\mnras},
         year = 2017,
        month = jan,
       volume = {464},
       number = {1},
        pages = {897-914},
          doi = {10.1093/mnras/stw2397},
archivePrefix = {arXiv},
       eprint = {1605.03462},
 primaryClass = {astro-ph.CO},
       adsurl = {https://ui.adsabs.harvard.edu/abs/2017MNRAS.464..897B}
}

@ARTICLE{Viel13wdm,
   author = {{Viel}, M. and {Becker}, G.~D. and {Bolton}, J.~S. and {Haehnelt}, M.~G.
	},
    title = "{Warm dark matter as a solution to the small scale crisis: New constraints from high redshift Lyman-{$\alpha$} forest data}",
  journal = {\prd},
archivePrefix = "arXiv",
   eprint = {1306.2314},
 primaryClass = "astro-ph.CO",
     year = 2013,
    month = aug,
   volume = 88,
   number = 4,
      eid = {043502},
    pages = {043502},
      doi = {10.1103/PhysRevD.88.043502},
   adsurl = {http://adsabs.harvard.edu/abs/2013PhRvD..88d3502V}
}

@ARTICLE{Rorai2017,
       author = {{Rorai}, Alberto and {Hennawi}, Joseph F. and {O{\~n}orbe}, Jose and {White}, Martin and {Prochaska}, J. Xavier and {Kulkarni}, Girish and {Walther}, Michael and {Luki{\'c}}, Zarija and {Lee}, Khee-Gan},
        title = "{Measurement of the small-scale structure of the intergalactic medium using close quasar pairs}",
      journal = {Science},
         year = 2017,
        month = apr,
       volume = {356},
       number = {6336},
        pages = {418-422},
          doi = {10.1126/science.aaf9346},
archivePrefix = {arXiv},
       eprint = {1704.08366},
 primaryClass = {astro-ph.CO},
       adsurl = {https://ui.adsabs.harvard.edu/abs/2017Sci...356..418R}
}

@BOOK{Peebles1980,
       author = {{Peebles}, P.~J.~E.},
        title = "{The large-scale structure of the universe}",
         year = 1980,
       adsurl = {https://ui.adsabs.harvard.edu/abs/1980lssu.book.....P}
}

@ARTICLE{Noaz2007,
       author = {{Naoz}, S. and {Barkana}, R.},
        title = "{The formation and gas content of high-redshift galaxies and minihaloes}",
      journal = {\mnras},
         year = 2007,
        month = may,
       volume = {377},
       number = {2},
        pages = {667-676},
          doi = {10.1111/j.1365-2966.2007.11636.x},
archivePrefix = {arXiv},
       eprint = {astro-ph/0612004},
 primaryClass = {astro-ph},
       adsurl = {https://ui.adsabs.harvard.edu/abs/2007MNRAS.377..667N}
}

@ARTICLE{Tseliakhovich2010,
       author = {{Tseliakhovich}, Dmitriy and {Hirata}, Christopher},
        title = "{Relative velocity of dark matter and baryonic fluids and the formation of the first structures}",
      journal = {\prd},
         year = 2010,
        month = oct,
       volume = {82},
       number = {8},
          eid = {083520},
        pages = {083520},
          doi = {10.1103/PhysRevD.82.083520},
archivePrefix = {arXiv},
       eprint = {1005.2416},
 primaryClass = {astro-ph.CO},
       adsurl = {https://ui.adsabs.harvard.edu/abs/2010PhRvD..82h3520T}
}

@ARTICLE{Barkana2001,
       author = {{Barkana}, R. and {Loeb}, A.},
        title = "{In the beginning: the first sources of light and the reionization of the universe}",
      journal = {\physrep},
         year = 2001,
        month = jul,
       volume = {349},
       number = {2},
        pages = {125-238},
          doi = {10.1016/S0370-1573(01)00019-9},
archivePrefix = {arXiv},
       eprint = {astro-ph/0010468},
 primaryClass = {astro-ph},
       adsurl = {https://ui.adsabs.harvard.edu/abs/2001PhR...349..125B}
}

@ARTICLE{Eisenstein2007,
       author = {{Eisenstein}, Daniel J. and {Seo}, Hee-Jong and {White}, Martin},
        title = "{On the Robustness of the Acoustic Scale in the Low-Redshift Clustering of Matter}",
      journal = {\apj},
         year = 2007,
        month = aug,
       volume = {664},
       number = {2},
        pages = {660-674},
          doi = {10.1086/518755},
archivePrefix = {arXiv},
       eprint = {astro-ph/0604361},
 primaryClass = {astro-ph},
       adsurl = {https://ui.adsabs.harvard.edu/abs/2007ApJ...664..660E}
}

@ARTICLE{Smith2008,
       author = {{Smith}, Robert E. and {Scoccimarro}, Rom{\'a}n and {Sheth}, Ravi K.},
        title = "{Motion of the acoustic peak in the correlation function}",
      journal = {\prd},
         year = 2008,
        month = feb,
       volume = {77},
       number = {4},
          eid = {043525},
        pages = {043525},
          doi = {10.1103/PhysRevD.77.043525},
archivePrefix = {arXiv},
       eprint = {astro-ph/0703620},
 primaryClass = {astro-ph},
       adsurl = {https://ui.adsabs.harvard.edu/abs/2008PhRvD..77d3525S}
}

@ARTICLE{Cain2024,
       author = {{Cain}, Christopher and {Scannapieco}, Evan and {McQuinn}, Matthew and {D'Aloisio}, Anson and {Trac}, Hy},
        title = "{The hydrodynamic response of small-scale structure to reionization drives large IGM temperature fluctuations that persist to z = 4}",
      journal = {\mnras},
         year = 2024,
        month = sep,
       volume = {533},
       number = {1},
        pages = {L100-L106},
          doi = {10.1093/mnrasl/slae067},
archivePrefix = {arXiv},
       eprint = {2405.02397},
 primaryClass = {astro-ph.CO},
       adsurl = {https://ui.adsabs.harvard.edu/abs/2024MNRAS.533L.100C}
}

@ARTICLE{Cain2026,
       author = {{Cain}, Christopher and {Das}, Aloha and {D'Aloisio}, Anson and {Foreman}, Simon and {Scannapieco}, Evan and {Moreno}, Esteban and {Lugatiman}, Matthew and {Cohon}, Joshua and {Maksora Tohfa}, Hurum and {Trac}, Hy},
        title = "{Introducing SAGUARO -- Simulating IGM Evolution and Environments At High Resolution: Setup and First Results}",
      journal = {arXiv e-prints},
         year = 2026,
        month = mar,
          eid = {arXiv:2603.25788},
        pages = {arXiv:2603.25788},
          doi = {10.48550/arXiv.2603.25788},
archivePrefix = {arXiv},
       eprint = {2603.25788},
 primaryClass = {astro-ph.CO},
       adsurl = {https://ui.adsabs.harvard.edu/abs/2026arXiv260325788C}
}

@ARTICLE{GarciaGallego2025,
       author = {{Garcia-Gallego}, Olga and {Ir{\v{s}}i{\v{c}}}, Vid and {Haehnelt}, Martin G. and {Viel}, Matteo and {Bolton}, James S.},
        title = "{Constraining mixed dark matter models with high-redshift Lyman-alpha forest data}",
      journal = {\prd},
         year = 2025,
        month = aug,
       volume = {112},
       number = {4},
          eid = {043502},
        pages = {043502},
          doi = {10.1103/4k29-h99l},
archivePrefix = {arXiv},
       eprint = {2504.06367},
 primaryClass = {astro-ph.CO},
       adsurl = {https://ui.adsabs.harvard.edu/abs/2025PhRvD.112d3502G}
}

@ARTICLE{GarciaGallego2025b,
       author = {{Garcia-Gallego}, Olga and {Ir{\v{s}}i{\v{c}}}, Vid and {Haehnelt}, Martin G. and {Bolton}, James S.},
        title = "{Constraints on the Thompson optical depth to the CMB from the Lyman-$α$ forest}",
      journal = {arXiv e-prints},
         year = 2025,
        month = sep,
          eid = {arXiv:2510.00107},
        pages = {arXiv:2510.00107},
          doi = {10.48550/arXiv.2510.00107},
archivePrefix = {arXiv},
       eprint = {2510.00107},
 primaryClass = {astro-ph.CO},
       adsurl = {https://ui.adsabs.harvard.edu/abs/2025arXiv251000107G}
}

@ARTICLE{Berg2025,
       author = {{Berg}, T. and {D'Odorico}, V. and {Boera}, E. and {Calderone}, G. and {Cuellar}, R. and {Cupani}, G. and {Cristiani}, S. and {Di Stefano}, S. and {Grazian}, A. and {Guarneri}, F. and {Ir{\v{s}}i{\v{c}}}, V. and {Lopez}, S. and {Milakovi{\'c}}, D. and {Noterdaeme}, P. and {Pasquini}, L. and {Viel}, M. and {Welsh}, L.},
        title = "{From the Intergalactic to the Interstellar Scales - EQUALS: a High-resolution Legacy Survey of Gas in the Distant Universe Using ESPRESSO}",
      journal = {The Messenger},
         year = 2025,
        month = sep,
       volume = {195},
        pages = {23-26},
          doi = {10.18727/0722-6691/5395},
archivePrefix = {arXiv},
       eprint = {2512.06159},
 primaryClass = {astro-ph.GA},
       adsurl = {https://ui.adsabs.harvard.edu/abs/2025Msngr.195...23B}
}

@ARTICLE{Artola2024,
       author = {{Artola}, Ander and {Bosman}, Sarah E.~I. and {Gaikwad}, Prakash and {Davies}, Frederick B. and {Nasir}, Fahad and {Farina}, Emanuele P. and {Protu{\v{s}}ov{\'a}}, Klaudia and {Puchwein}, Ewald and {Spina}, Benedetta},
        title = "{Signatures of warm dark matter in the cosmological density fields extracted using Machine Learning}",
      journal = {arXiv e-prints},
         year = 2024,
        month = nov,
          eid = {arXiv:2411.17853},
        pages = {arXiv:2411.17853},
          doi = {10.48550/arXiv.2411.17853},
archivePrefix = {arXiv},
       eprint = {2411.17853},
 primaryClass = {astro-ph.CO},
       adsurl = {https://ui.adsabs.harvard.edu/abs/2024arXiv241117853A}
}

@ARTICLE{Kaiser1987,
       author = {{Kaiser}, Nick},
        title = "{Clustering in real space and in redshift space}",
      journal = {\mnras},
         year = 1987,
        month = jul,
       volume = {227},
        pages = {1-21},
          doi = {10.1093/mnras/227.1.1},
       adsurl = {https://ui.adsabs.harvard.edu/abs/1987MNRAS.227....1K}
}

@ARTICLE{Paranjape2026,
       author = {{Paranjape}, Aseem and {Sheth}, Ravi K.},
        title = "{Impact of fiducial cosmology in model-agnostic cosmological inference with the BAO feature}",
      journal = {arXiv e-prints},
         year = 2026,
        month = jun,
          eid = {arXiv:2606.06591},
        pages = {arXiv:2606.06591},
          doi = {10.48550/arXiv.2606.06591},
archivePrefix = {arXiv},
       eprint = {2606.06591},
 primaryClass = {astro-ph.CO},
       adsurl = {https://ui.adsabs.harvard.edu/abs/2026arXiv260606591P}
}

@ARTICLE{Becker2015,
       author = {{Becker}, George D. and {Bolton}, James S. and {Lidz}, Adam},
        title = "{Reionisation and High-Redshift Galaxies: The View from Quasar Absorption Lines}",
      journal = {\pasa},
         year = 2015,
        month = dec,
       volume = {32},
          eid = {e045},
        pages = {e045},
          doi = {10.1017/pasa.2015.45},
archivePrefix = {arXiv},
       eprint = {1510.03368},
 primaryClass = {astro-ph.CO},
       adsurl = {https://ui.adsabs.harvard.edu/abs/2015PASA...32...45B}
}

@ARTICLE{Eilers2017,
       author = {{Eilers}, Anna-Christina and {Davies}, Frederick B. and {Hennawi}, Joseph F. and {Prochaska}, J. Xavier and {Luki{\'c}}, Zarija and {Mazzucchelli}, Chiara},
        title = "{Implications of z {\ensuremath{\sim}} 6 Quasar Proximity Zones for the Epoch of Reionization and Quasar Lifetimes}",
      journal = {\apj},
         year = 2017,
        month = may,
       volume = {840},
       number = {1},
          eid = {24},
        pages = {24},
          doi = {10.3847/1538-4357/aa6c60},
archivePrefix = {arXiv},
       eprint = {1703.02539},
 primaryClass = {astro-ph.GA},
       adsurl = {https://ui.adsabs.harvard.edu/abs/2017ApJ...840...24E}
}

@ARTICLE{Keating2024,
       author = {{Keating}, Laura C. and {Bolton}, James S. and {Cullen}, Fergus and {Haehnelt}, Martin G. and {Puchwein}, Ewald and {Kulkarni}, Girish},
        title = "{JWST observations of galaxy-damping wings during reionization interpreted with cosmological simulations}",
      journal = {\mnras},
         year = 2024,
        month = aug,
       volume = {532},
       number = {2},
        pages = {1646-1658},
          doi = {10.1093/mnras/stae1530},
archivePrefix = {arXiv},
       eprint = {2308.05800},
 primaryClass = {astro-ph.GA},
       adsurl = {https://ui.adsabs.harvard.edu/abs/2024MNRAS.532.1646K}
}

@ARTICLE{Ma2026,
       author = {{Ma}, Ke and {Bolton}, James S. and {Ir{\v{s}}i{\v{c}}}, Vid and {Gaikwad}, Prakash and {Puchwein}, Ewald},
        title = "{An improved model for the effect of correlated Si III absorption on the one-dimensional Lyman-{\ensuremath{\alpha}} forest power spectrum}",
      journal = {\mnras},
         year = 2026,
        month = feb,
       volume = {546},
       number = {1},
          eid = {staf2262},
        pages = {staf2262},
          doi = {10.1093/mnras/staf2262},
archivePrefix = {arXiv},
       eprint = {2509.08613},
 primaryClass = {astro-ph.CO},
       adsurl = {https://ui.adsabs.harvard.edu/abs/2026MNRAS.546f2262M}
}

@ARTICLE{DAloisio2020,
       author = {{D'Aloisio}, Anson and {McQuinn}, Matthew and {Trac}, Hy and {Cain}, Christopher and {Mesinger}, Andrei},
        title = "{Hydrodynamic Response of the Intergalactic Medium to Reionization}",
      journal = {\apj},
         year = 2020,
        month = aug,
       volume = {898},
       number = {2},
          eid = {149},
        pages = {149},
          doi = {10.3847/1538-4357/ab9f2f},
archivePrefix = {arXiv},
       eprint = {2002.02467},
 primaryClass = {astro-ph.CO},
       adsurl = {https://ui.adsabs.harvard.edu/abs/2020ApJ...898..149D}
}

@ARTICLE{Nasir2020,
       author = {{Nasir}, Fahad and {D'Aloisio}, Anson},
        title = "{Observing the tail of reionization: neutral islands in the z = 5.5 Lyman-{\ensuremath{\alpha}} forest}",
      journal = {\mnras},
         year = 2020,
        month = may,
       volume = {494},
       number = {3},
        pages = {3080-3094},
          doi = {10.1093/mnras/staa894},
archivePrefix = {arXiv},
       eprint = {1910.03570},
 primaryClass = {astro-ph.CO},
       adsurl = {https://ui.adsabs.harvard.edu/abs/2020MNRAS.494.3080N}
}

@ARTICLE{Hirata2018,
       author = {{Hirata}, Christopher M.},
        title = "{Small-scale structure and the Lyman-{\ensuremath{\alpha}} forest baryon acoustic oscillation feature}",
      journal = {\mnras},
         year = 2018,
        month = feb,
       volume = {474},
       number = {2},
        pages = {2173-2193},
          doi = {10.1093/mnras/stx2854},
archivePrefix = {arXiv},
       eprint = {1707.03358},
 primaryClass = {astro-ph.CO},
       adsurl = {https://ui.adsabs.harvard.edu/abs/2018MNRAS.474.2173H}
}

%%%%%%%%%%%%%%%%%%%%%%%%%%%%%%%%%%%%%%%%%%%%%%%%%%

%%%%%%%%%%%%%%%%% APPENDICES %%%%%%%%%%%%%%%%%%%%%

\appendix
\section*{Appendices}

\section{Numerical convergence}
\label{appendix:0}

\begin{figure*}
\centering
\includegraphics[width=0.95\textwidth]{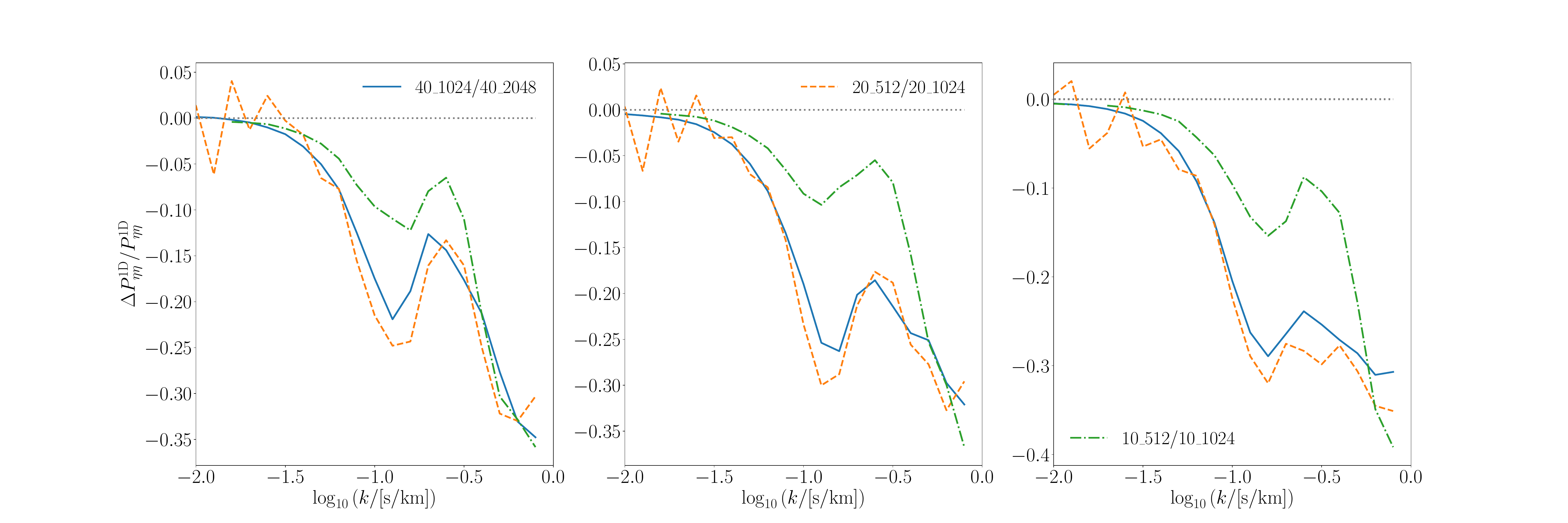}
\includegraphics[width=0.95\textwidth]{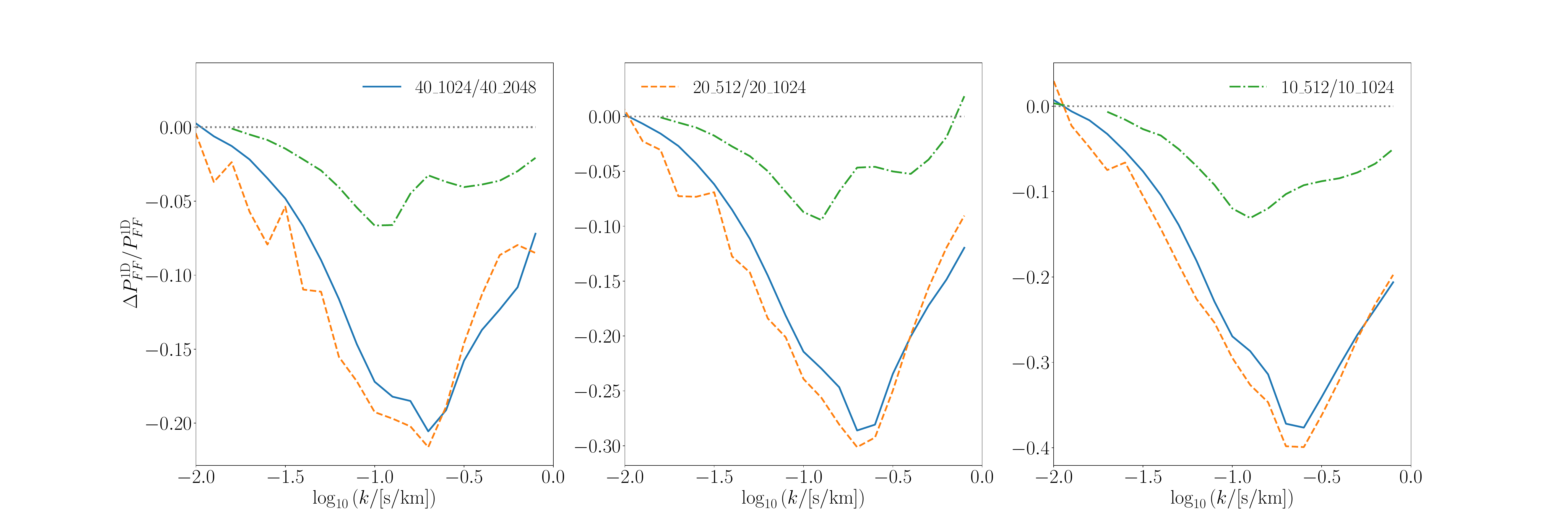}
\caption{The numerical convergence tests using simulations with different box sizes and number of particles. The three panels in each row are for $z=4.2, 4.6$ and $5.0$ from left to right respectively. The numerical convergence corrections range from lower resolution simulations (blue solid; orange dashed) to high resolution simulations (green dot-dashed). The difference between $40$ and $20\,h^{-1}\mathrm{cMpc}$ box size at fixed resolution is highlighted between (blue solid) and (orange dashed) models. {\it Top:} The ratio of the lower resolution to higher resolution $P_{\eta\eta}$ power spectrum, showing general stability of the peak position with the numerical convergence of the simulation. {\it Bottom:} Repeat of the results from \citep{Irsic2024} for the resolution correction ratio for $P_{F}(k)$, extended to smaller scales. Similarly to the velocity power, the position of the small-scale feature is largely preserved.
}
\label{fig:numerical_convergence}
\end{figure*}

The limitations of numerical convergence of cosmological simulations are determined by two main factors -- box size which typically limits the number of large-scale modes; and the mass or particle resolution that limits the smallest resolved scales. Fig.~\ref{fig:numerical_convergence} shows the effects of the resolution correction at fixed box sizes, which is the more important correction for small-scale physics. The bottom row shows the results that have been previously known in the literature (e.g. \cite{Irsic2024}), on the flux power spectrum resolution correction changes with scale and redshift. Generally the correction is larger at higher redshifts, and more important on smaller scales. As already noted in \citep{Puchwein2023,Irsic2024} the resolution correction in the flux power spectrum has a scale-dependent feature on small-scales that is highly dependent on the thermal history. The position of this feature indicates that the thermal history dependence on the resolution correction is tightly linked to the peculiar velocity field, suggesting an interplay between gas density, temperature and peculiar velocity having different effective smoothing scales at a fixed particle numbers in a hydrodynamical simulation.

The effects are similar in the velocity power spectra, where the effect is stronger at higher redshifts, and shows a strong scale dependence at small-scales, with the amplitude of the resolution correction reaching  amplitudes of 20-40\%. The exact scale dependence around the small scale peak of Fig.~\ref{fig:models} in $P_{\eta\eta}$ is also imprinted on the resolution correction for the field, suggesting similar origins.

\begin{figure*}
\centering
\includegraphics[width=0.95\textwidth]{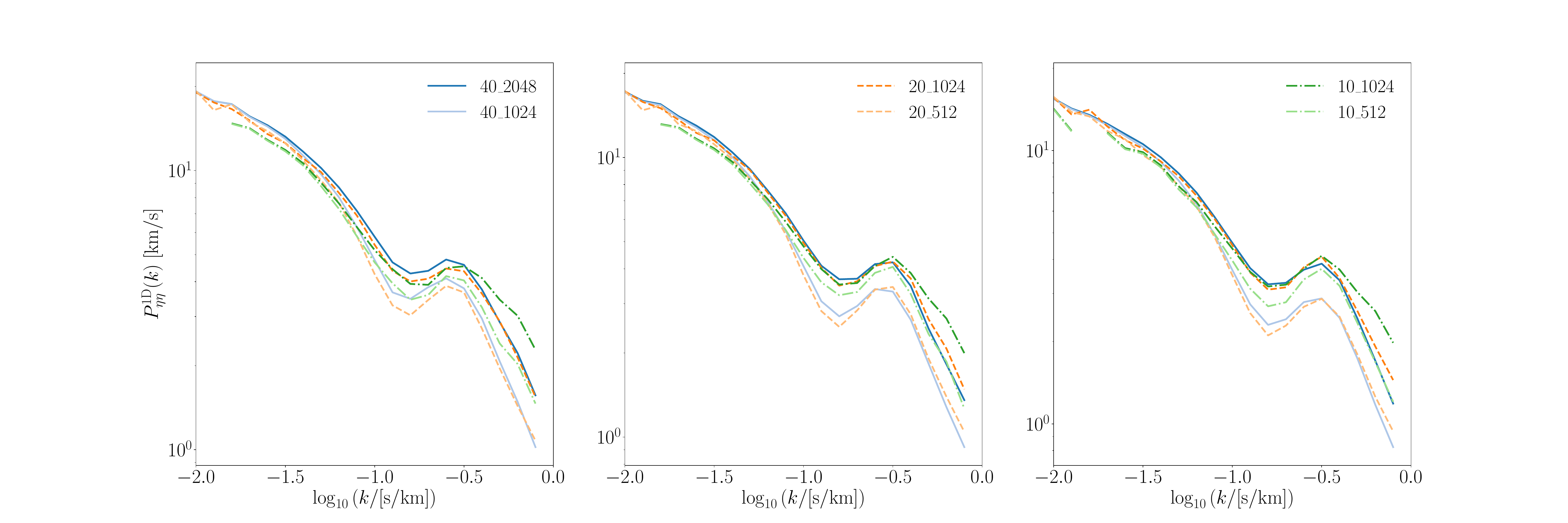}
\caption{The numerical convergence tests using simulations with different box sizes and number of particles. The three panels in each row are for $z=4.2, 4.6$ and $5.0$ from left to right respectively. The numerical convergence corrections range from lower resolution simulations (blue solid; orange dashed) to high resolution simulations (green dot-dashed). The difference between $40$ and $20\,h^{-1}\mathrm{cMpc}$ box size at fixed resolution is higlighted between (blue solid) and (orange dashed) models. The figure focuses on the effect of different box size and particle number choices directly on $P_{\eta\eta}$ field, showcasing their effect on the small-scale feature imprinted by the pressure smoothing scale.
}
\label{fig:numerical_convergence_v2}
\end{figure*}

This effect however is not completely responsible for the feature in either flux or velocity power spectrum that is being treated as a signal in this study. As shown in the top panels of Fig.~\ref{fig:numerical_convergence_v2}, the effect of different box sizes and particle numbers on $P_{\eta\eta}$ predominantly changes the amplitude around the small-scale peak feature, but leaves the position generally intact. The effect of higher resolution does however broaden the peak structure, especially at low redshifts. This suggests that in higher resolution simulations better capture the non-linear effects in the peculiar velocity field that lead to non-linear damping of the acoustic features.

\section{Spherical toy model}
\label{appendix:A}

The density and velocity profiles of each expanding sphere are modelled with a spherically symmetric radial profile, following a Gaussian (for density) and first order Hermite (for velocity). The Gaussian model of the density profile imposes a linear expansion with time for the standard deviation parameter, $R(t) = R + v_e (t-t_0)$, with a fixed expansion velocity $v_e$ and initial size of the sphere characterized by $R=R(t_0)$. The amplitude of the profile changes with time according to mass conservation law, which results in the amplitude of the density profile decaying with time $\propto \left(R/R(t) \right)^3$. 

The continuity equation in spherical coordinates and under the symmetry of the problem, is used to derive the velocity profile ($u_r$) of the expanding sphere as
\begin{align}
    \frac{1}{r^2} \frac{\partial}{\partial r}\left[ r^2 \rho(r,t) u_r(r,t) \right] &=  - \frac{\partial \rho}{\partial t}\\
    \rho(r,t) &= \rho_0 \left(\frac{R}{R(t)}\right)^3 e^{-\frac{r^2}{2R^2(t)}},
\end{align}
where $\rho(r,t)$ and $u_r(r,t)$ are the density and radial velocity profiles. In the context of redshift-space distortion effect ($\eta$), the important quantity will be the density weighted radial velocity $v_r(r,t) = (1+\delta) u_r(r,t)$, where the density contrast is $1+\delta = \rho(r,t)/\langle \rho(r,t) \rangle_r$, and defined with respect to the radially averaged density profile $\langle \rho(r,t) \rangle_r$. This leads to the solution for the density weighted velocity profile as
\begin{align}
    v_r(r,t) = \left[1+\delta(r,t)\right]\,u_r(r,t) & = \frac{v_e}{\sqrt{2\pi}} \frac{r}{R(t)} e^{-r^2/2R(t)^2},\\
    v\equiv v_r(R) &= \frac{v_e}{\sqrt{2 e\pi}}
\end{align}
where the peak velocity of the profile $v=v_r(R(t))$ has been defined as a parameter of the model.

In the spherically symmetric case, the 3D radial velocity profile only has a component in the radial direction of each individual sphere. For a given set of lines of sight through the simulation volume, impact parameter from each sphere to a given line of sight will give a contribution to the projected velocity profile imprinted on that sightline direction. However, in practice, spheres with an impact parameter value greater than $3\times R(t)$ do not contribute significantly to the velocity field along a given line of sight.

This is the toy model used in Section.~\ref{sec:toy_model}. The model parameters are thus the size of the sphere $R(t)$, the peak velocity of the radial profile $v=v_r(R(t))$ and the number density of such spheres put in the simulation box. The values of $R(t)$ and $v$ were chosen to be fixed for all spheres. A more physical picture would include variations of $R,v$ with spatial position within the simulation box, reflecting that certain parts reionized earlier than others. Similarly, the spheres could include a clustering component beyond a spatially homogeneous number density. However, these alterations do not change the conclusions of the model and its applicability to explain the small-scale signal in $\eta$ and flux fields.

The model showcases how the redshift-space component of the expanding spheres can be used to model the signal in simulations. Even though mathematically elegant, the model, as any toy model, comes with caveats. In particular, it lacks the external pressure that would arise from spheres expanding {\it into} the surrounding gas. If the surrounding gas is at some background density $\rho_b$ then the solution to the radial velocity profile $v_r$ is modified to include a term of $\rho(r,t)/(\rho_b + \rho(r,t))$. Although this changes the Hermitian shape and thus the rate at which the radial profile decreases, it does not change the results significantly, at the expense of an additional parameter to the model that captures the ratio of the density of the sphere and the background density $\rho_0/\rho_b$.

All components of the toy model were constructed in the comoving coordinates.

\section{Empirical fitting functions}
\label{appendix:B}

This appendix provides details of the empirical fitting functions used to extract the physical size of the pressure smoothing scale from the simulations in both velocity and flux power spectrum space.

The velocity power spectra are moved into the comoving distance space, where the velocities of the simulations are naturaly defined. Moreover, the amplitude of the velocity power spectra is re-scaled by the linear theory expectation that argues that on large scales the velocity gradient is proportional to the density perturbation, $\eta_k \propto (faH) \delta_k$. In linear theory, all of the redshift dependence is encoded in the pre-factor that depends on the linear growth rate ($f={\rm d}\ln{D}/{\rm d}\ln{a}$) and the conformal Hubble expansion rate ($aH$). The remaining redshift evolution of the peak position in the velocity power spectrum is fitted with an empirical model of the form
\begin{align}
    (f a \frac{H}{H_0})^2 P_{\eta\eta}^{\rm 1D} &= P_{\rm smooth}^\eta(k) + P_{\rm peak}^\eta(k), \\
    P_{\rm smooth}^\eta(k) &= A^2 k^n e^{-\frac{k^2}{k_s^2}}, \notag \\
    P_{\rm peak}^\eta(k) &= B^2 \left(\frac{k}{k_p}\right)^2 e^{-\frac{k^2}{k_p^2}} + C^2\left(\frac{k}{k_p}\right)^2 \frac{1}{\left(1+\frac{k^2}{k_p^2}\right)^{7/4}}. \notag
    \label{eq:eta}
\end{align}

The fitting is split into three components, one describing the overall smooth baseline level of the velocity power spectra, that sets the power at large scales and the subsequent fall-off of the power towards small-scales beyond a typical scale $k_s$; and two components describing the peak. There is only one scale associated with the peak position ($k_p$), and the two components describe the shape of the peak and overall decay of power on scales much smaller than the peak position. The model has six free parameters, aside from the two scales ($k_s$,$k_p$) for the smooth and peak components, respectively, the model also allows for variations of the amplitudes of all three components, and for the power-law behaviour of the smooth component in the low-k regime. In particular the last component guarantees that at $k \gg k_p > k_s$ the velocity power spectrum falls off as $\propto k^{-3/2}$ as observed in simulations. All the components of the fit are empirical in nature.

Similarly, for the purpose of fitting the peak position in the flux space, the flux power spectrum amplitude has been re-scaled. Typical result of simulations (e.g. \citep{bolton17}) is that the amplitude of the flux power spectrum scales roughly with the mean transmission squared, or a square of the effective optical depth ($\tau_{\rm eff} = -\ln{\langle F \rangle}$). However, as is well known this also introduces an apparent transfer of power between large and small scales, and means that models at different redshift cross one another. This introduced numerical instability in our fitting procedure, and we have thus opted for a re-scaling that is less physically motivated, but that preserves the unique scalings for each of the simulation outputs. For that reason we rescale flux power spectrum amplitude by the inverse of the effective optical depth.

The empirical model fitted to the flux power spectra has the form of 
\begin{align}
    \tau_{\rm eff}^{-1} P_{FF}^{\rm 1D} &= P_{\rm smooth}^F(k) + P_{\rm peak}^F(k), \\
    P_{\rm smooth}^F(k) &= A^2 \left( \frac{k}{k_s}\right)^{-1/2} \frac{1}{\left(1 + (k/k_s)^b\right)^c}, \notag \\
    P_{\rm peak}^F(k) &= B^2 \exp{\left[-\left(\frac{k}{k_p}\right)^{3/2}-\left(\frac{k}{k_p}\right)^{1/2}\right]}, \notag
    \label{eq:flux}
\end{align}
with power-law index $b$ numerically given by $b = 7.17/c$. 

As before, the fitting for the flux power spectrum, is split into smooth and peak components, each with associated scale. The model has five free parameters, aside from the two scales ($k_s$,$k_p$), also the amplitudes of the smooth and peak component, as well as a power-law behaviour of the smooth component in the high-k regime. This power-law scaling of the smooth component becomes important at all redshifts at scales $k>>k_p$, and it dominates the signal at low redshifts.

In the case the velocity power spectrum, both peak components determine the exact position of the peak in the velocity power spectrum, while the smooth component does not contribute significantly. Whereas for the flux power spectra the peak position is determined as a local maximum of the ratio of the peak and smooth components, and the exact peak position depends also on the parameters of the smooth component. We find these choices to well reproduce the simulation results when estimating the peak position directly from the simulated power spectra. The advantage of the fitting procedure is to minimize the scatter associated with finding a peak in the finite k-binned function, as well as to properly propagate the uncertainty on the fitting parameters to the final peak positions.

% Don't change these lines
\bsp	% typesetting comment
\label{lastpage}
\end{document}